\documentclass[superscriptaddress,twocolumn,showpacs,preprintnumbers,amsmath,amssymb,pra]{revtex4-1}

\usepackage{graphicx}
\usepackage[english] {babel}
\usepackage{booktabs}
\usepackage{color}

\begin{document}

\title{Time-dependent generalized-active-space configuration-interaction approach to photoionization 
dynamics of atoms and molecules}

\author{S.~Bauch}
  \author {L.K.~S\o{}rensen}
\affiliation{%
Department of Physics and Astronomy, 
Aarhus University,  Aarhus 8000 C, Denmark
}%

 \affiliation{%
Institut f\"ur Theoretische Physik und Astrophysik, Christian-Albrechts-Universit\"at zu Kiel, Leibnizstra\ss{}e 15, 24098 Kiel, Germany
}%

\author{L.B.~Madsen}
\affiliation{%
Department of Physics and Astronomy, 
Aarhus University,  Aarhus 8000 C, Denmark
}%

\date{\today}

\begin{abstract}
We present a wave-function based method to solve the time-dependent many-electron 
Schr\"odinger  equation (TDSE) with special emphasis
on strong-field ionization phenomena. The theory builds 
on the configuration-interaction (CI) approach supplemented by the 
generalized-active-space (GAS) concept from quantum chemistry. The latter
allows for a controllable reduction in the number of 
configurations in the CI expansion by imposing restrictions on the active orbital space. 
The method is similar to the recently
formulated time-dependent restricted-active-space (TD-RAS) CI method 
[D. Hochstuhl, and M. Bonitz, Phys. Rev. A {\bf 86}, 053424 (2012)].
We present details of our implementation and address convergence properties 
with respect to the active spaces and the associated account of electron correlation in
both ground state and excitation scenarios.  
We apply the TD-GASCI theory to strong-field 
ionization of polar diatomic molecules and illustrate how the method allows us to uncover 
a strong correlation-induced shift of the 
preferred direction of emission of photoelectrons.

\end{abstract}



\maketitle

\section{Introduction}
Tracing electron motion and correlation on their natural time scales has become possible
within the last decade due to enormous experimental progress in light-pulse technology and 
detection methods~\cite{brabec2000,krausz2009,schuette2012,bauch2012}.
These experimental advancements and associated new possibilities 
for elucidating quantum motion on an ultrafast timescale challenge theory.
Clearly, approaches that treat electron correlation and at the same time are 
explicitly time-dependent are needed to fully exploit the potential of the experimental 
capabilities. 
The development and application of such a time-dependent (TD) 
quantum theory for the many-electron problem (MEP) including a possibly strong external 
field is the topic  of the present work.

Over the years, various approaches for the solution of the TDMEP on a quantum-mechanical 
level have been  proposed and applied. On the one hand, there exist approximative methods 
which consider a reduced number of electrons (typically one or two) in precalculated 
pseudo potentials created by frozen electrons that are assumed to be
inactive in the considered dynamics, apart from contributing to the potential governing the
motion of the active electrons. This approach results in the single- and 
two-active-electron approximations (SAE/TAE) for which the time-dependent Schr\"odinger 
equation (TDSE) has been solved for photoionization, high-order harmonic 
generation (HHG) and related phenomena since the  late 80's~\cite{kulander1987}. 
The appeal of these methods is their flexibility and numerical feasibility with respect to 
the considered systems. 
The dynamical effects of the frozen electrons, however, 
cannot be tested within these SAE and TAE approximations, and likewise there is no
 explicit account of electron 
correlation in general.
To this end, approaches have been developed where all electrons  
are treated simultaneously on different levels of ``activity''. Numerical tractable
methods are either achieved by approximating the electron-electron  (e-e)
interactions or by reducing the configuration space.
Among these methods are the time-dependent configuration-interaction (TD-CI)
method  and its 
truncations, where in particular the simplest TD-CI-singles (TD-CIS) with only single-orbital excitation out of the Hartree-Fock (HF) ground state 
has been applied
\cite{Krause2005,Rohringer2006}.
In addition, we mention  the
time-dependent density functional theory~\cite{wilken2007},
time-dependent natural orbital theory~\cite{brics2013}, 
time-dependent coupled-cluster theory~\cite{kvaal2012}, the non-equilibrium Green's functions approaches
~\cite{balzer2010,balzer2010c,hochstuhl2010},
and the state-specific expansion approach~\cite{schultze2010,mercouris2010}.
Up to now,  in particular the time-dependent 
R-Matrix theory~\cite{hart2007,lysaght2008,lysaght2009,hart2014}
and the 
multi-configurational time-dependent Hartree Fock (MCTDHF) 
method~\cite{hochstuhl2010,hochstuhl2011,caillat2005,nest2005,kato2004,meyer1990,Haxto2012}
have found applications in the photoionization community.
In the perturbative regime for the matter-light interaction, the MCTDHF method has been
applied to the determination of inner-shell photoionization cross sections for molecular 
hydrogen fluoride~\cite{Haxto2012}. The number of configurations in the 
MCTDHF method increases exponentially with respect to the number of electrons
due to the full-CI expansion. This makes the method infeasible for systems having more than a few electrons interacting with a strong field.
The TD complete-active-space self-consistent-field method (TD-CASCF)~\cite{sato2013}, 
and the more general TD restricted-active space 
SCF (TD-RASSCF)~\cite{miyagi2013,miyagi2014,miyagi2014b} cure this scaling 
by imposing restrictions on the active orbital spaces, while keeping the attractive  SCF
notion of the MCTDHF approach, i.e., the orbitals are time-dependent and  optimally updated 
in each time-step.

In this paper, we consider the TD generalized-active-space (GAS) CI concept, which 
is based on a general CI truncation scheme adapted from (time-independent) quantum 
chemistry. In the GAS/RAS approach~\cite{olsen1988,fleig2001} the single-particle basis is partitioned into physically motivated subsets and only the configurations 
that are expected to be most relevant for the processes under consideration
are included in the CI expansion, and thus reducing the number of 
configurations considerably.
By specifying the GAS, generalizations of the SAE and TAE approximations,
without the need of contracting pseudo potentials, are readily obtained as limiting cases.
Moreover CI truncations, such as CIS, CIS-doubles (CISD), CISD-triples (CISDT), etc. 
can be easily specified and the method, accordingly, allows 
a straightforward increase in the account of electron correlation within a
specified active orbital space.
The present method is similar to the time-dependent restricted-active-space 
(TD-RAS) CI scheme~\cite{hochstuhl2012}, which was applied to calculate the photoionization cross sections of Beryllium and Neon~\cite{hochstuhl2013}.

A fundamental problem of any truncated CI method is the choice of
a good orbital basis. In this work, we address this issue with the focus on time-dependent 
excitations and give a detailed analysis of different choices: 
pseudo-orbitals based on HF orbitals similar to~\cite{hochstuhl2012}, an adapted version for larger systems
and natural orbitals.
Further, we demonstrate in the limiting case of 4 electrons
the convergence of the method by detailed comparison with fully-correlated 
TDSE  or equivalent calculations.
In addition, we give details of the implementation and extend the approach to small molecules in strong
external fields. In particular the approach allows us to uncover a strong effect of 
electron correlation on the preferred emission direction of photoelectrons. 

The paper is organized as follows. Section~\ref{sec:theory} outlines 
the concepts of CI and GAS and introduces the equations of
motion and notations used in this work. In Sec.~\ref{sec:mixed-basis},
we address the problem of photoionization and the related
choice of appropriate orbital basis sets. 
Here, we choose a partially rotated basis, which combines orbital and grid-based
approaches in an efficient manner.
In Sec.~\ref{sec:numerical}, we apply the TD-GASCI method to model systems for atomic 
helium and beryllium and compare with
fully correlated results.
We especially focus on convergence properties with respect to the GAS partitions
and the choice of orbitals. Our analysis covers ground state properties as well as excitation 
scenarios.
Finally, the application of TD-GASCI  is extended to 
molecular systems. We focus on the polar
diatomic lithium hydride (LiH) molecule. 
After a discussion of its ground state properties, we present a study of the strong-field
ionization with single-cycle laser pulses including electron correlation effects.
Section~\ref{sec:conclusions} summarizes and concludes.

\section{Theory}
\label{sec:theory}
We aim to provide a general scheme for the numerical treatment of the
non-relativistic 
many-electron time-dependent Schr\"odinger equation (ME-TDSE) which is 
particularly well-suited for the description of ionization processes
of atoms and molecules by short and/or strong pulses. 

The fundamental equation is the TDSE for $N_{\textup{el}}$ electrons 
in an atom or a molecule with fixed nuclei
(atomic units are used throughout),
\begin{equation}
 i \frac{\partial}{\partial t} |\Psi\ (t) \rangle = \hat{H}(t) |\Psi (t) \rangle \;,
 \label{eq:tdse}
 \end{equation}
with the time-dependent Hamiltonian
\begin{equation}
 \hat{H}(t)=\sum_{i=1}^{N_{\textup{el}}} \hat{h}_i(t) + \sum_{i<j}^{N_{\textup{el}}} \hat{w}_{ij} \;.
 \label{eq:td-hamil}
\end{equation}
The single-particle term referring to particle $i$,
\begin{equation}
 \hat{h}_i(t)=\hat{t}_i+\hat{v}_i+\boldsymbol{F}(t)\hat{\boldsymbol{r}}_i,
 \label{eq:sp-hamil}
\end{equation}
consists of the kinetic energy $\hat{t}_i$, the potential describing the 
attractive interaction with the nuclei $\hat{v}_i$, and the interaction 
with the external field, $\boldsymbol{F}(t)\hat{\boldsymbol{r}}_i$. The latter being described 
in the dipole approximation
within the length gauge.
The two-body part of $\hat{H}(t)$ is given by the binary interaction between electrons $i$ and $j$, $\hat{w}_{ij}$.

The general solution of Eq.~\eqref{eq:tdse} is only feasible by employing powerful numerical techniques. Pioneering work in the
context of (strong-field) ionization was done for (effective) one-electron systems in Refs.~\cite{krause1992,kulander1987,kulander1991,schafer1993}.
For systems with interacting electrons, only very few cases are manageable without approximations, such as helium
and H$_2$~\cite{taylor2005,parker2006,nepstad2010,simonsen2012,laulan2003,colgan2002,pazourek2012,feist2008,feist2009}, and even in these cases
the whole range of laser frequencies and intensities can not be accessed.

When the number of electrons increases only approximate solutions are accessible, 
see, e.g., Ref.~\cite{hochstuhl2014} for a thorough review, 
and it is mandatory to go beyond the level of time-dependent Hartree-Fock (TDHF) to allow for a description of electron-electron 
correlation effects.

\subsection{Time-dependent Configuration-Interaction}
Let us form Slater determinants $|\Phi_ I\rangle$
from the spin orbitals $\vert \phi_i\rangle = \vert \varphi_i \rangle \otimes \vert \sigma \rangle $ to 
construct  the many-electron basis.  Here $\vert \sigma \rangle $ with 
$\sigma = \{ \alpha,\beta\} $ denotes the spin degree of freedom, and
$\vert \varphi_i \rangle$ the remaining single-particle degrees of freedom.
The multi-index $I$
specifies the individual configurations spanning the full CI 
Fock space ${\cal V}_\text{FCI}$.
The expansion of $|\Psi (t) \rangle$ into this 
basis set with time-dependent coefficients $C_I(t)$,
\begin{equation}
|\Psi(t)\rangle=\sum_{I \in {\cal V}_{\text{FCI}} }C_I(t) | \Phi_I\rangle \;,
\label{eq:slater-expansion}
\end{equation}
gives the matrix form of the TDSE,
\begin{equation}
 i\frac{\partial}{\partial t} C_I(t) = \sum_{J \in {\cal V}_\text{FCI}}   H_{IJ}(t) C_J(t) \;,
 \label{tdse-matrix}
\end{equation}
with $H_{IJ}(t)=\langle \Phi_I |\hat{H}(t) | \Phi_J \rangle$.
The matrix representation of $\hat{H}(t)$ is referred  to as the CI-matrix in the following.

The CI-matrix elements are conveniently determined using the 
language of second quantization. In  the occupation number representation 
$|\boldsymbol{n}\rangle$ and $|\boldsymbol{m}\rangle$,
the matrix element are then given by~\cite{helgaker}  
\begin{eqnarray}
 \langle \boldsymbol{n}|\hat{H}|\boldsymbol{m}\rangle & =& \sum_{pq}h_{pq}(t) \sum_\sigma \langle \boldsymbol{n}| \hat{c}_{p\sigma}^\dagger \hat{c}_{q\sigma} |\boldsymbol{m} \rangle +\nonumber \\
 & &+ \frac{1}{2} \sum_{pqrs} w_{pqrs} \sum_{\sigma\tau}\langle \boldsymbol{n} |\hat{c}_{p\sigma}^\dagger \hat{c}_{r\tau}^\dagger \hat{c}_{s\tau} \hat{c}_{q\sigma} |\boldsymbol{m}\rangle \;,
 \label{eq:hmatrix-elements}
\end{eqnarray}
where a spin-free Hamiltonian, i.e., the same spatial orbital for $\alpha$ and $\beta$ spin is assumed, 
and where $\hat{c}_{p\sigma}$ ($\hat{c}_{p\sigma}^{\dagger}$) denotes the  annihilation (creation) operator of the spin-orbital $|\varphi_p\rangle \otimes | \sigma \rangle$.
Here, the \emph{one-electron} integrals
\begin{equation}
 h_{pq}=t_{pq}+v_{pq}(t) \; ,
\end{equation}
of the kinetic and potential energy contributions to the single-particle part are given by
\begin{eqnarray}
 t_{pq}&=&-\frac{1}{2}\int \textup{d}\boldsymbol{r} \; \varphi_p^*(\boldsymbol{r}) \nabla^2 \varphi_q(\boldsymbol{r})\;, \nonumber \\
 v_{pq}(t)&=& \int \textup{d}\boldsymbol{r} \; \varphi_p^*(\boldsymbol{r}) v(\boldsymbol{r};t) \varphi_q(\boldsymbol{r})\;,
 \label{eq:oneel}
\end{eqnarray}
with $v(\boldsymbol{r};t)=v(\boldsymbol{r})+\boldsymbol{F}(t)\boldsymbol{r}$,
and the \emph{two-electron} integrals of the interaction by (note that we use the chemist's notation of the integrals \cite{helgaker})
\begin{eqnarray}
 w_{pqrs}=\;\;\;\;\;\;\;\; \nonumber\\
 \iint \textup{d}\boldsymbol{r}_1\textup{d}\boldsymbol{r}_2 \; \varphi_p^*(\boldsymbol{r}_1)\varphi_r^*(\boldsymbol{r}_2)w(\boldsymbol{r}_1,\boldsymbol{r}_2)\varphi_q(\boldsymbol{r}_1) \varphi_s(\boldsymbol{r}_2)\;.
 \label{eq:twoel}
 \end{eqnarray}
 Especially the nature of the two-electron integrals~\eqref{eq:twoel} imposes practical restrictions on the
underlying single-particle basis since for general basis sets the number of matrix elements scales as 
$\mathcal{O}(N_b^4)$ with $N_b$ being the number of spatial orbitals $\varphi_i(\boldsymbol{r})$ 
[corresponding to $2N_b$ spin orbitals $\phi_i(z)$, $z=(\boldsymbol{r}, \sigma)$]. A way to cure this unfavorable scaling in the context of 
photoionization-related problems, which involves the electronic continuum and hence 
necessarily a large $N_b$, is described in Sec.~\ref{sec:mixed-basis}.

Up to this point, Eq.~\eqref{tdse-matrix} is exact and inherits the full complexity of the MEP,
and the approach is referred to as full CI (FCI).
The number of configurations $n_{\textup{conf}}$ [or number of Slater determinants in Eq.~\eqref{eq:slater-expansion}] spanning ${\cal V}_\text{FCI}$ 
scales as
\begin{equation}
 n_{\textup{conf}}={2 N_b \choose N_{\textup{el}}} \;.
 \label{eq:exp-conf}
\end{equation}
In principle, a reduction by some factor by exploiting symmetries of the system, such as spin and spatial symmetries,
is possible~\cite{hochstuhl2014}. 
In the following, we will assume conservation of the total spin
for our spin-independent Hamiltonian~(\ref{eq:td-hamil})-(\ref{eq:sp-hamil}).
Still FCI calculations are only feasible for a very limited number 
of spin orbitals $2N_b$ and few electrons~\cite{olsen1990,gan2005} and therefore mostly used 
to benchmark other approximative methods.

To overcome this fundamental barrier, the CI expansion~\eqref{eq:slater-expansion} has to be truncated at a certain level.
Frequently used are CIS, CISD and so on, in which one takes into account only
singly, doubly or higher excited determinants with the hope to capture the dominant correlation 
contributions. Especially in the context of photoionization
and related phenomena, the truncation at the singles level has some tradition~\cite{Krause2005,gordon2006,Rohringer2006,Krause2007,Krause2009,Rohringer2009,pabst2013,karamatskou2014,greenman2010,Luppi2012},
since photoionization into a structureless continuum can often be described accurately
in a single-electron picture.

In this work, we take a more general approach by partitioning the single-particle basis into physically motivated subsets and choosing determinants
that are expected to be most relevant for the processes under consideration. 
This concept is known as generalized (or restricted) active space (GAS/RAS)
in the quantum chemistry literature~\cite{olsen1988,fleig2001}.
A time-dependent realization based on a time-independent spin-orbital basis was presented in Ref.~\cite{hochstuhl2012}, and in an SCF setting in 
Refs.~\cite{miyagi2013,miyagi2014,miyagi2014b}.
The idea of selecting determinants by their importance, and thus truncating the 
CI expansion, has a long tradition in atomic and molecular
physics~\cite{cremer2012}. 

\subsection{GAS scheme}
\label{ssec:gas-scheme}

\begin{figure}
 \includegraphics[width=6cm]{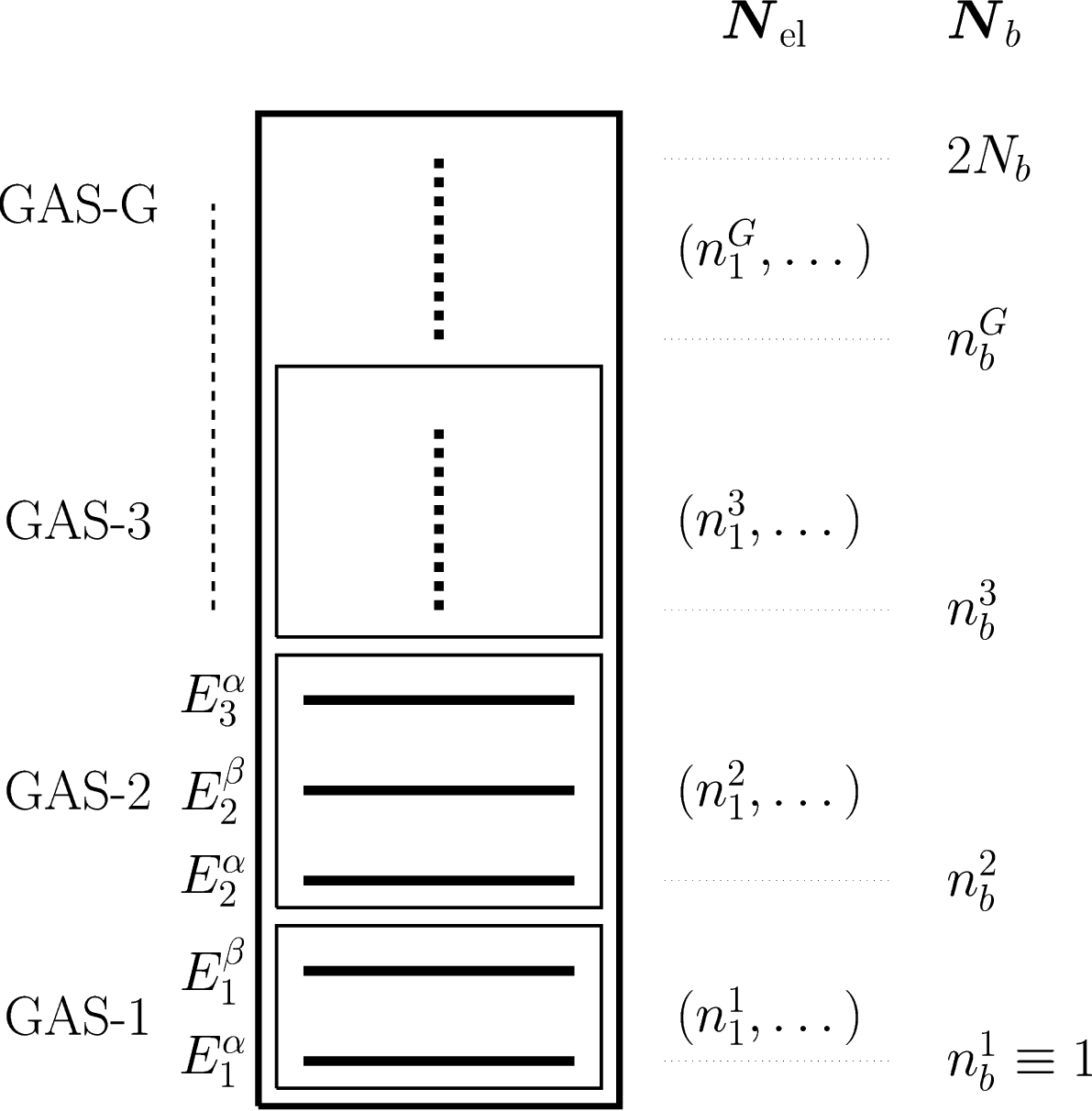}
 \caption{Schematic of the generalized-active-space (GAS) method with $G$ subspaces GAS-1 to GAS-G. 
 The spin-orbital partition is given by $\boldsymbol{N}_b=[1,n_b^2,\dots]$
 and the allowed number of electrons in each subspace by $\boldsymbol{N}_{\textup{el}}=[(n_1^1,\dots),\dots,(n_1^G,\dots)]$. 
 The energy eigenvalues of the single-particle orbitals are labeled by $E_1^{\alpha,\beta}$, where $\alpha$ and $\beta$ denote
 the spin coordinate. For the non-relativistic studies in this work, 
 these are degenerate, $E_i^\alpha=E_i^\beta$.}
 \label{fig:gas-schematic}
\end{figure}

The configuration space is determined by two arrays of numbers. The first array, 
$\boldsymbol{N}_b$,
contains information about  the partition of the single-particle 
spin-orbital basis into the $G$ subspaces of the GAS. 
We may order the single-particle basis in any desired way. For the
present discussion it is convenient to assume that the spin-orbitals
are ordered according to their energy. The lowest orbital is
indexed by 1, the next (possibly degenerate) by 2, etc until the highest-lying spin-orbital, 
which is indexed by $2 N_b$, the total number of spin-orbitals.
The notation $n_b^1=1$ means that subspace 1 in the GAS partitioning 
contains spin-orbitals from the lowest one, 1.
Then $n_b^2$ denotes the value of the spin-orbital index for the lowest-lying
spin-orbital in the second subspace, $n_b^3$ the   index of the lowest-lying 
spin-orbital in the third subspace, and 
so forth [Fig.~\ref{fig:gas-schematic}]. This information is summarized in 
$\boldsymbol{N}_b$, containing 
the string of indices
 \begin{equation}
\boldsymbol{N}_b=[n_b^1\equiv1,n_b^2,\dots,n_b^G]\;.
\end{equation}
The second array specifies the number of occupied spin-orbitals that we allow in each subspace of the GAS partitioning,
\begin{equation}
\boldsymbol{N}_{\textup{el}}=[(n^1_1,n^1_2,\dots),\dots,(n^G_1,n^G_2,\dots)] \;.
\end{equation}
As an illustrative, but not practical relevant example, consider a single-particle 
basis with only 6 spin-orbitals corresponding
to 3 different spatial orbitals and 3 different energies for a two-electron system,
which are degenerate w.r.t. spin projection.
Let $G=2$, such that we have 2 active subspaces denoted by GAS-1 and GAS-2. Assume
we choose the first subspace 
to include only the two lowest degenerate spin-orbitals, and the 
second to include the remaining four. In this case 
 $\boldsymbol{N}_b = [n_b^1=1,n_b^2= 3]$. The specification of 
 $\boldsymbol{N}_{\textup{el}}$ determines the amount of correlation that is 
 taken into account between these orbitals.
 For example,  we 
 could consider  $\boldsymbol{N}_{\textup{el}}= [(n_1^1=2,n_2^1=1), 
 (n_1^2=0,n_2^2= 1)]$, which allows 2 or 1 occupied orbital in GAS-1 and 0 or 1
 occupied orbital in GAS-2.
The set of occupation numbers with subscript 1,   i.e., the combination 
$[(n_1^1=2), (n_1^2=0)]$
 corresponds to configurations with  both lowest-lying  spin-orbitals
  occupied in the lowest subspace, GAS-1, and no occupied spin-orbitals in the other subspace, GAS-2. 
The other set of occupation numbers with subscript 2, i.e., the 
  combination $[(n_2^1=1), (n_2^2=1)]$, describes 
 one-particle excitation out of GAS-1 into GAS-2. In both 
 cases $\sum_j^G n_i^j=N_\text{el}$ for all $i$ as it should be. If we had chosen
 $\boldsymbol{N}_{\textup{el}}= [(n_1^1=2,n_2^1=1,n_3^1=0), 
 (n_1^2=0,n_2^2= 1,n_2^2=2)]$ we would have included double excitation out 
 of GAS-1 (doubles) in 
 addition to the singles of the previous example. 
 It is clear that such partitioning  
in the general case allows  the realization of any 
excitation scheme. It is also clear that introduction of restrictions on the 
excitation between the different GASs dramatically 
reduce $n_\text{conf}$.

In the context of this work, we focus mainly on excitation phenomena with one-electron 
continua, i.e., excitations, where we allow one electron to be removed from 
the bound-state part of the spectrum, described, for example, by the 
GAS-1, GAS-2, GAS-3 
in Fig.~\ref{fig:gas-schematic} and excited to the GAS describing the 
continuum, GAS-G. To relate to the commonly used notation in
quantum chemistry, we denote this case by
CAS$^*(N_{\textup{el}}^C,K)$. Here
CAS refers to ``Complete-Active-Space'', 
$N_{\textup{el}}^C$ denotes the number of electrons in the active space, and
 $K$ the number of single-particle spatial orbitals in the active space.
Finally, the star indicates
that single excitations out of the active space have been added
compared to the  usual  CAS scheme (sometimes also written as $[N_{\textup{el}}^C,K]$-CAS~\cite{jensen2007}). 
\begin{figure}
 \includegraphics[width=8.2cm]{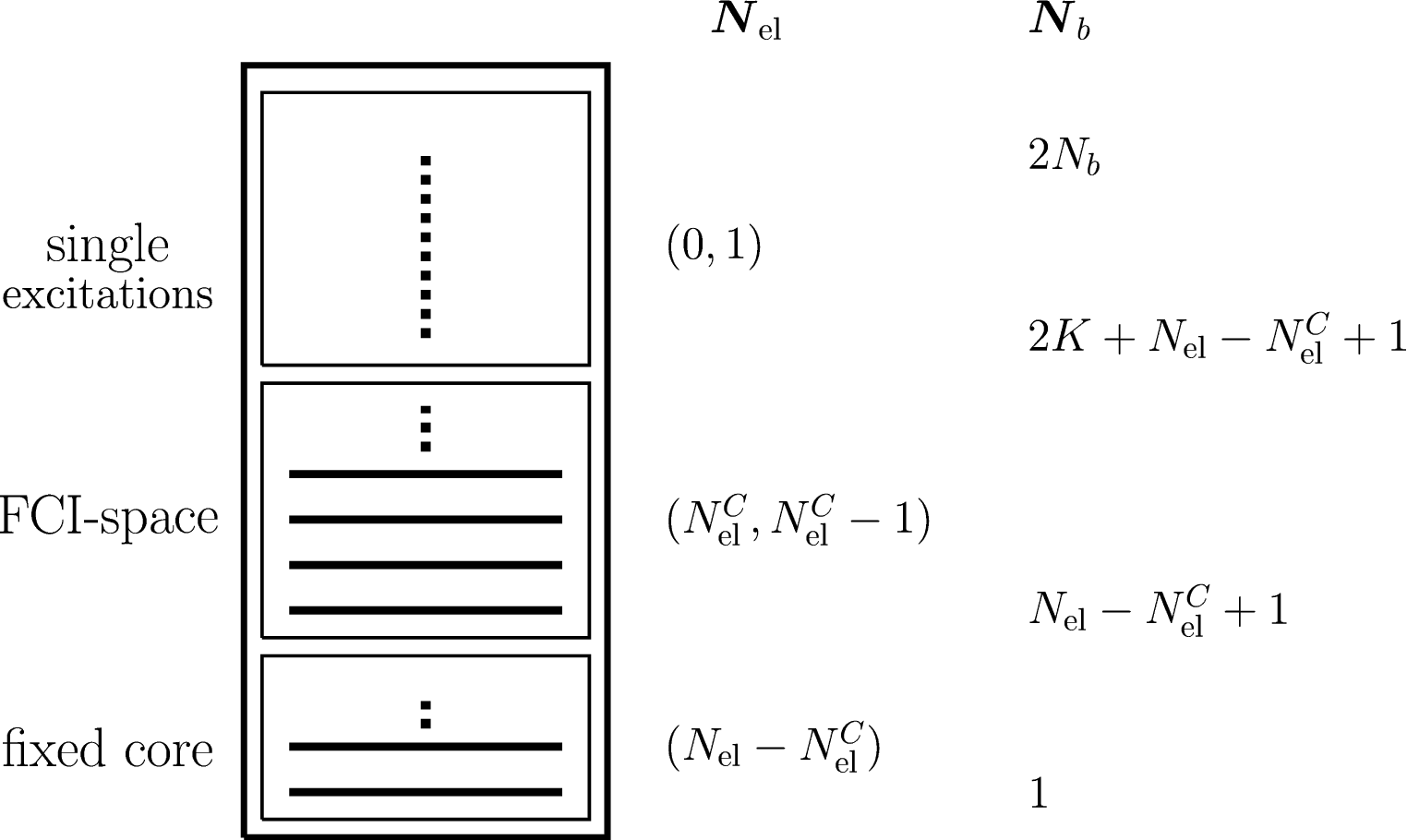}
 \caption{Schematic of the GAS partitioning mainly used in this work.
 The scheme is denoted by CAS$^*(N_{\textup{el}}^C,K)$ and consists of a fixed
 core with $N_{\textup{el}}-N_{\textup{el}}^C$ electrons in the same number
 of spin orbitals (this space is empty if $N_\textup{el}^C=N_{\textup{el}}$) and
 an active space with $N_\textup{el}^C$ electrons in $2K$ spin orbitals from
 which one electron can be removed and excited into GAS-G describing the 
 one-electron continuum.
 }
 \label{fig:cas-schematic}
\end{figure}

The CAS$^*$ scheme is illustrated in Fig.~\ref{fig:cas-schematic}. The lowest GAS-1 describes a fixed core with
$N_\textup{el}-N_{\textup{el}}^C$ electrons, where each electron occupies one spin orbital. This space is empty if one chooses to include all electrons in the active space, 
corresponding to the specification CAS$^*(N_\textup{el},K)$. GAS-2 is 
the active space with $N_\textup{el}^C$
electrons occupying $2K$ spin orbitals, for which all possible configurations are constructed, and in this sense  a FCI description is maintained in this space.
On top of that, we allow for single excitations from GAS-2 to GAS-3, i.e., we remove one electron from GAS-2,
resulting in $N_{\textup{el}}^C-1$ electrons in GAS-2, and create it in GAS-3.
The number of electrons in the individual subspaces and the 
corresponding partition of the single-particle spin-orbital basis are given in Fig.~\ref{fig:cas-schematic}, right columns.
Although some of the electrons may be kept frozen within the GAS scheme, i.e., some spin orbitals
are always occupied, we emphasize that
their interaction potential with all other electrons contributes to the sum in the Hamiltonian~\eqref{eq:hmatrix-elements}
and no pseudo potentials for the explicitly active electrons need to be set up.

Using the GAS concept, the CI expansion~\eqref{eq:slater-expansion} reduces in size,
\begin{equation}
  |\Psi^\textup{GAS} (t)\rangle = \sum_{I \in {\cal V}_\textup {GAS}} C_I(t) |\Phi_I\rangle \;,
  \label{eq:gas-expansion}
\end{equation}
where only configurations within the specified Fock space ${\cal V}_\text{GAS}$ contribute. The corresponding set of differential 
equations for the amplitudes reads
\begin{equation}
 i\frac{\partial}{\partial t} C_I(t) = \sum_{J \in {\cal V}_\text{GAS}}   
 H^{\textup{GAS}}_{IJ}(t) C_J(t) \;.
\label{eq:tdse-matrix-gas}
\end{equation}
All limiting cases for CI calculations, such as SAE, CIS, CISD etc., up
to FCI can be realized by the appropriate GAS scheme~\cite{hochstuhl2012}.

The solution of Eq.~\eqref{eq:tdse-matrix-gas} requires a choice of a
 single-particle spin-orbital basis
$| \varphi_i \rangle \otimes |\sigma \rangle $, which allows for an efficient GAS expansion in 
terms of Slater determinants.
Once the single-particle basis is constructed and the one- and two-electron integrals, Eqs.~\eqref{eq:oneel} and \eqref{eq:twoel},
are evaluated, the GASCI matrix $H^{\textup{GAS}}_{IJ}(t)$ can be calculated. A straightforward way to evaluate 
Eqs.~\eqref{eq:oneel} and \eqref{eq:twoel},
is by applying  Slater-Condon 
rules~\cite{helgaker}, but this approach is 
in practice limited to a rather small
determinantal space due to the high degree of sparsity of the Hamiltonian and the 
unavoidable ``calculation'' of zero-elements in $H^{\textup{GAS}}_{IJ}(t)$. 
An alternative efficient way 
already proposed in the 80's in the original formulation of RAS-CI~\cite{olsen1988}
overcomes the latter problem
by decomposing the excitations into $\alpha$ and $\beta$ spin strings and employing a lexicographical 
ordering of the determinants. 
This approach was also taken in Ref.~\cite{hochstuhl2012},
and variations thereof in Refs.~\cite{miyagi2013,miyagi2014,miyagi2014b}.
In this work, we use a generalized scheme based on the 
construction and manipulation of \emph{types}
of excitation classes which is particularly suited for GAS calculations and 
which has previously been successfully applied in Coupled-Cluster theory~\cite{soerensen2010,soerensen2011}.
In this approach the  zero parts of  the CI matrix are identified and omitted 
from the calculation and only the remaining non-zero blocks are calculated 
and stored in a sparse matrix format.
Besides, the scheme offers a very efficient way of setting up the CI matrix with a minimal number of evaluation of the 
electron integrals and provides a strategy for parallelization.
Additional information and a detailed description of the reformulated integral direct method and 
algorithm is to be found in a forthcoming publication~\cite{soerensen2014}.

\subsection{Time propagation}
To solve Eq.~\eqref{eq:tdse-matrix-gas}, we first set up $H^{\textup{GAS}}_{IJ}(t)$.
The solution of  Eq.~\eqref{eq:tdse-matrix-gas} is given by discretization of the time variable $t=N_t\Delta t$ into $N_t$ time steps and successive application of
the time evolution operator $U(t,t+\Delta t)=\exp\left[-i H(t+\Delta t) \Delta t \right]$ to the vector of coefficients $C_I(t)$. 
In order to avoid a diagonalization of the (large) CI matrix $H^{\textup{GAS}}_{IJ}(t)$ 
at each time step, we employ an Arnoldi-Lanczos procedure and propagate the matrix equation
in the corresponding Krylov subspace (we typically use a Krylov dimension of $10$), which results in a unitary and stable
propagation scheme. Details of the time-propagation algorithm can be found in Refs.~\cite{park1986,beck2000}.
This method involves only matrix multiplications of $H^{\textup{GAS}}_{IJ}(t)$ with the coefficient vector $C_I(t)$ often referred to as ``$\sigma$-vector-step'' in the CI literature~\cite{fleig2001}, and which can be performed efficiently using sparse matrix algebra and by 
exploiting block structures of the CI-matrix.
The initial condition $| \Psi^{\textup{GAS}}(t=0) \rangle \equiv | \Psi^{\textup{GAS}}_0\rangle$ for Eq.~\eqref{eq:gas-expansion} [or $C_I(t=0)\equiv C_I^0$ for Eq.~\eqref{eq:tdse-matrix-gas}]
is prepared through imaginary time propagation (ITP)  
by replacing $t \rightarrow \imath t$,  see, 
e.g.,~Refs.\cite{kosloff1986,bauer2006}. To obtain the correctly correlated initial
state, it is crucial to use exactly the same parameters with respect to the single-particle
basis and the GASCI scheme as in the real time propagation.

\section{Basis sets}
\label{sec:mixed-basis}
In this section, we discuss the spatial part of the single-particle basis functions.
For the convergence of truncated CI expansions, the choice of the single-particle basis 
plays a crucial role. Roughly speaking, 
the single-particle basis used to form the Slater determinants for the many-particle basis 
should closely resemble the physical one- and many-electron
excitations of the system. For ground state CI calculations, it can be shown that the CI 
expansion converges fastest using natural orbitals~\cite{szabo}. The most common 
approach 
is the use of HF reference states or improved orbitals which incorporate
part of the e-e correlation contribution on the single-particle level.
However, all of these orbital-based expansions with good properties for the 
ground- and bound-state CI expansions 
become essentially inapplicable in the limit of spatially extended systems.
This is caused by the highly non-favorable scaling of the two-electron integrals with the number of single-particle basis functions,
$\mathcal{O}(N_b^4)$.

\subsection{Partially rotated basis}
In order to allow for photoionization processes with large computational grids,
we follow a different strategy~\cite{hochstuhl2012} and use a partially rotated~\cite{basis-remark} basis set.
In the following, we will work out the formulas for the one-dimensional (1D) case. 
Analogous expressions  in 3D spherical coordinates~\cite{hochstuhl2014} or
prolate spheroidal coordinates~\cite{tao2009} are straightforward and pose
no conceptual difficulties. In short, the technique can be summarized by using localized HF-like orbitals for the 
description of the bound part of the spectrum and a grid-like representation for the continuum part. A similar
technique was developed in Ref.~\cite{yip2011}.

\begin{figure}
 \includegraphics[width=8.5cm]{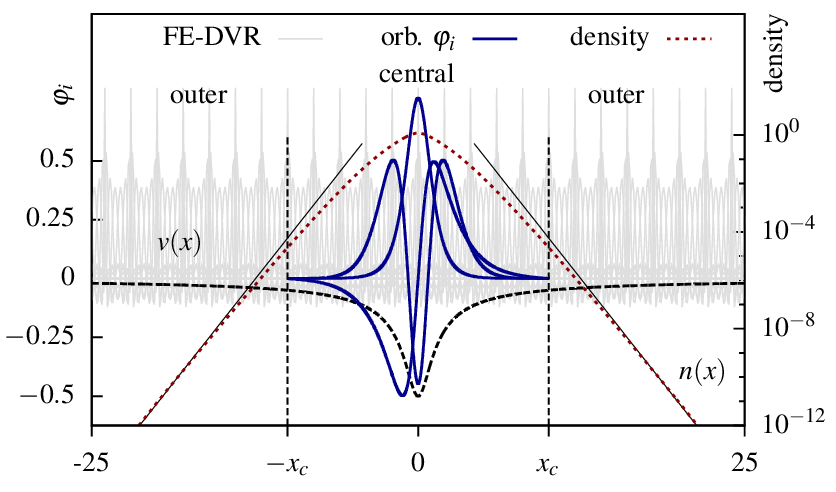}
 \caption{(color online). Schematic view of the partially rotated basis set with pseudo orbitals 
 $\varphi_i$ in the central region $[-x_c,x_c]$ (solid, blue lines)
 close to the minimum of the binding potential (dashed, black line). 
  In addition, a correlated single-particle density, cf. Eq.~\eqref{eq:sp-density-1d},
  for the ground state of a model for beryllium (see Sec.~\ref{ssec:1dbe})
 calculated for the complete computational grid $[-x_s,x_s]$ is given by the dashed, red line (logarithmic scale to the right). 
 The asymptotics is indicated by the thin black lines labeled ``$n(x)$'', cf.~Eq.~\eqref{eq:asymp-den}.
 The underlying FE-DVR grid (all functions are included in the calculation)
 is sketched in gray. Parameters in the Figure are $x_s=30$ and $x_c=10$.
 We used 30 elements with 8 DVR functions per element. Quantities on the abscissa are given in atomic units (a.u.).}
  \label{fig:mixed-basis}
\end{figure}

Let us consider a single-particle basis composed of finite-element discrete-variable representation (FE-DVR) functions~\cite{rescigno2000}. 
Similar expressions and strategies can be developed, e.g., with 
B-splines~\cite{bachau2001}.
The FE-DVR basis consists of $N_e$ elements, which discretize the simulation box ranging from $[-x_s,x_s]$
into partitions 
\begin{equation}
-x_s=x^0<\dots<x^i<x^{i+1}<\dots<x^{N_e}=x_s\;.
\label{eq:fedvr-partition}
\end{equation}
Each element, $[x^i,x^{i+1}]$, is spanned by $n_g$ DVR functions.
The basis functions are given by (we follow the notation in Refs.~\cite{balzer2010,balzer2010b})
\begin{eqnarray}
 \chi_m^i&=& \frac{f^i_{n_g-1}(x)+f^{i+1}_0(x)}{\sqrt{w^i_{n_g-1}+w^{i+1}_0}} \;\;\; \textup{for} \;\; m=0 \;\;\textup{(bridge)} 
 \label{eq:fedvr-bridge} \\
 \chi_m^i&=& \frac{f_m^i(x)}{\sqrt{w_m^i}} \;\;\;\hspace{2cm} \textup{else}\;\; \textup{(element)}\;,
 \label{eq:fedvr-element} 
\end{eqnarray}
with the Lobatto shape functions
\begin{equation}
 f_m^i(x)=\prod_{\bar{m}\neq m} \frac{x-x_{\bar{m}}^i}{x_m^i-x_{\bar{m}}^i}
 \label{eq:lobatto}
\end{equation}
and the Gau\ss{}-Lobatto quadrature points $x_m^i$ and weights $w_m^i$. The lower index $m$
labels the DVR function and the upper index $i$ the corresponding element.
The bridge functions~\eqref{eq:fedvr-bridge} connect adjacent elements $i$ and $i+1$ and assure communication between both
elements and the continuity of the wave function.
The overall basis is schematically drawn in Fig.~\ref{fig:mixed-basis} with gray lines. 
The bridge functions~\eqref{eq:fedvr-bridge} have spiky shape.

The FE-DVR matrix elements 	of Eqs.(\ref{eq:oneel})-(\ref{eq:twoel}) 
possess a simple form~\cite{balzer2010,rescigno2000,schneider2006}:
\begin{eqnarray}
 v_{pq}&=&v(x_p)\delta_{pq} 
 \label{eq:fedvr-v} \\
 w_{pqrs}&=&w(x_p,x_r)\delta_{pq} \delta_{rs} \; ,
 \label{eq:fedvr-w} \\
 t_{pq}&\equiv& t_{m_1m_2}^{i_1i_2}  
 \label{eq:fedvr-t}\\
 & = & \frac{\left(\delta_{i_1i_2}+\delta_{i_1i_2\pm1}\right)}{2}\int \textup{d}x \left(\frac{\textup{d}}{\textup{d}x}\chi^{i_1}_{m_1}(x)\right)\left(\frac{\textup{d}}{\textup{d}x} \chi^{i_2}_{m_2}(x)\right),\nonumber
\end{eqnarray}
where we combined element indices $i$ and DVR function indices $m$ to multi-indices
$p,q,r,s$, in analogy to Eqs.~(\ref{eq:oneel})-(\ref{eq:twoel}).
We point out that the number of non-vanishing two-electron integrals~\eqref{eq:fedvr-w}
scales as $\mathcal{O}(N_b^2)$, and not $\mathcal{O}(N_b^4)$ as for arbitrary sets, 
which is of high practical importance for the present approach.

Unfortunately, a single Slater determinant constructed directly from the FE-DVR 
functions represents a poor reference state 
for the CI expansion. We therefore follow Ref.~\cite{hochstuhl2012} 
and partition the basis set
interval $[-x_s,x_s]$ into a central part $[-x_c,x_c]$
close to the nuclei and a remaining outer part for
$|x|>x_c$, cf. Fig.~\ref{fig:mixed-basis}. The partition point $x_c$ is chosen such that it coincides with an 
FEDVR element partition $x^i$ between elements $i$ and $i+1$.
The FE-DVR basis set is thus partitioned as
\begin{eqnarray}
\chi^\textup{c}_p(x)&\equiv& \chi^j_m(x) \;\; \forall \;\; j,m : x^j_m\in(-x_c,x_c)  \;\;\; \textup{[central]}\;, \nonumber \\
\chi^\textup{o}_p(x)&\equiv& \chi^j_m(x) \;\; \textup{else} \;\; \textup{[outer]} \; .
\label{eq:basis-oc}
\end{eqnarray}
Note that 
the division of space into an inner and outer region, is also  central 
in (time-dependent)
R-Matrix theory~\cite{lysaght2008,lysaght2009,moore2011,hart2014,hart2007}.

In the following $\varphi_k(x)$ [$\varphi_k^{\textup{o}}(x)$] 
denotes an orbital localized in the central (outer) region. 
For $x\in(-x_c,x_c)$, orbitals (solid, blue  lines in Fig.~\ref{fig:mixed-basis})
with good reference properties, such as HF orbitals, are constructed, cf.~Sec.~\ref{ssec:orbitals}.
In terms of the FE-DVR functions, these are expressed as
\begin{eqnarray}
\varphi_k(x)&=&\sum_l b^{\textup{c}}_{lk} \chi^\textup{c}_l(x) \;\;\; \textup{with}\\
b^{\textup{c}}_{lk}&=&\int_{-x_c}^{x_c} \varphi_k(x)\chi_l^\textup{c}(x) \textup{d}x \;.
\label{eq:hf-trafo}
\end{eqnarray}
By excluding the bridge functions connecting the central with 
the outer region $|x|>x_c$ from the basis set in $(-x_c,x_c)$,
all orbitals are zero at $|x|=x_c$ by construction, i.e., $\varphi_k(x)\equiv 0$ for $|x|>x_c$.
In particular, $\varphi_i(x) \perp \varphi_j(x)$ and $\varphi_i(x) \perp \varphi_p^\textup{o}(x)=\chi_p^{\textup{o}}(x)$ holds.
The matrix elements are thus transformed by the matrix $\boldsymbol{b}^\textup{c}$
from Eq.~\eqref{eq:hf-trafo}.

Returning to the whole grid of $[-x_s,x_s]$, i.e., including all functions $\varphi^\textup{o}$ and the bridge functions at $\pm x_c$ into the basis set,
this transformation is continued such that the outer part remains unchanged,
\begin{equation}
 \boldsymbol{b}=\left(
 \begin{array}{ccc}
  \boldsymbol{1} & & \\
                 &\boldsymbol{b}^\textup{c} & \\
                 &                          & \boldsymbol{1}
  \end{array}
\right ) \;.  
\label{eq:transformb}
\end{equation}
The upper left corner corresponds to $x<-x_c$, the lower right to $x>x_c$. In practice, it is beneficial to
sort the basis such that the central part $\boldsymbol{b}^c$ is in the upper left corner, cf.~App.~\ref{ssec:app-transform}.

Using the unitary transformation~\eqref{eq:transformb} leaves the wave function unchanged 
(see App.~\ref{ssec:app-transform}).
Exploiting the $\delta$-structures of the FE-DVR matrix elements Eqs.~(\ref{eq:fedvr-v}-\ref{eq:fedvr-t}), very efficient scaling properties of the transformed integrals are obtained. Details of
the calculation and the storage scheme for one- and two-electron integrals
are given in App.~\ref{sec:app-electron-integrals}.
This approach allows for an accurate treatment of e-e interactions 
based on the CI expansion including  well-chosen single-particle basis functions 
close to the nuclei
as well as an efficient description  of wave packets in the continuum through the 
outer FE-DVR grid. 
We point out that
in contrast to the R-matrix approach, no special attention is needed for
assuring physical properties of the wave function across $\pm x_c$ separating the central 
and outer regions. The communication between the regions
is automatically assured by the bridge functions~\eqref{eq:fedvr-bridge}, which are constructed from the
Lobatto points at $|x|=x_c$ of the underlying FE-DVR basis set.

To demonstrate the smoothness of the wave function at the connection points after the basis transformation
and that the density has the correct asymptotic form, 
we show the single-particle density of the ground
state of a model for beryllium (see Sec.~\ref{ssec:1dbe}) after ITP
of the TDSE in Fig.~\ref{fig:mixed-basis}, (red) dashed line.
No ``jumps'' or discontinuities can be found, especially not at $\pm x_c$ and the density decays smoothly
over the whole simulation grid (note the logarithmic scale of the right axis).
The figure confirms that the asymptotic form of the density is 
\begin{equation}
n(x) \propto N \exp(-2 \varkappa | x | ) 
\label{eq:asymp-den}
\end{equation}
 with the parameter 
$\varkappa$ determined by $I_p = \varkappa^2/2$, $I_p$ the first ionization 
potential, and $N$ a proportionality constant. 
As a remark, we note that the well-known Brillouin theorem, which states that singly-excited determinants do
not lower the HF ground state energy, holds only for the central part $(-x_c,x_c)$ for our scheme. 
Increasing the grid to $[-x_s,x_s]$ and relaxing the GAS-CI wave function on the 
whole space lowers the ground state energy also if only single excitations into the
non-rotated part of the basis are included.
Illustrative examples are discussed in Secs.~\ref{sssec:1dhe-ge}
and \ref{sssec:1dbe-ge}.

\subsection{On the choice of single-particle orbitals in the central region}
\label{ssec:orbitals}
The central region, situated close to the nucleus $(-x_c,x_c)$, is 
described within a bound-state orbital basis set.
In Ref.~\cite{hochstuhl2012}  occupied HF orbitals and 
pseudo orbitals $\varphi_i(x)\equiv\varphi_i^{p_1}(x)$ constructed
from the interaction-free Hamiltonian $\hat{h}_0=\hat{t}+\hat{v}$ were used, with $\hat{v}$
the Coulomb attraction with the nucleus. The pseudo orbitals 
 are obtained from the eigenvalue problem
\begin{equation}
 \left(\hat{t} + \hat{v} \right)\varphi_i^{p_1} (x)= E^{p_1}_i \varphi_i^{p_1}(x)
 \label{eq:interaction-free-EVP}
\end{equation}
and a subsequent orthonormalization onto the occupied HF orbitals. 
They give an improved description of the virtual, i.e., unoccupied orbitals, 
compared to the virtual HF orbitals, which tend to be too delocalized [Fig.~4].
It turns out, however, that for situations with $N_{\textup{el}}>2$,
these hydrogen-like orbitals are strongly confined to the nucleus and 
do not describe valence orbitals well. 
This defect could possibly explain 
convergence issues related with photoionization of neon in Ref.~\cite{hochstuhl2013}.

One way around would be the use of an effective charge of the nucleus or a corresponding
quantum defect. However, this approach would need proper adjustments according to the
considered target. In order to obtain a flexible theory,
we propose to use generalized orbitals $\varphi_i^{p2}$,
which are defined by
\begin{equation}
 \left( \hat{t}+\hat{v}+\hat{v}_{\textup{H}}^{N_{\textup{el}}-2} \right) \varphi_i^{p_2}(x)=E_i^{p_2} \varphi_i^{p_2}(x)\;,
 \label{eq:hn2el}
\end{equation}
with the Hartree potential of the $N_{el}-2$ system, $\hat{v}_H^{N_\textup{el}-2}$.
In coordinate space, it has the form
\begin{equation}
v_H^{N_{\textup{el}-2}}(x)=\int dx' n^{\textup{HF}-2}(x') w(x,x')
\end{equation}
with the single-particle density $n^{\textup{HF}-2}(x)$ [see also Eq. (33)] which is obtained from a HF iteration with $N_\textup{el}-2$ electrons.
A subsequent orthogonalization of these orbitals to the occupied $N_{\textup{el}}$ HF orbitals (from a different HF calculation with $N_{\textup{el}}$ electrons) gives the
improved pseudo orbitals $\varphi_i^{p2}(x)$.

This choice is guided by physical intuition as for systems with one electron in the continuum (relevant for this study), a second electron in the vicinity 
to the nucleus moves in the effective potential of the $N_{\textup{el}}-2$ remaining electrons. We point out that for two-electron systems the 
choice of orbitals of non-interacting electrons, cf. Eq.~\eqref{eq:interaction-free-EVP}, coincides
with our improved version ($\hat{v}_{\textup{H}}^{N_{\textup{el}}-2}\equiv 0$ for two electrons).

As a third type of orbitals, we construct the natural orbitals, $\varphi_i^\textup{n}(x)$, 
which incorporate e-e correlation effects on the single-particle level.
They are constructed by first calculating the single-particle density matrix $\rho_{pq}$
within the central region from a highly-accurate GASCI (or, if possible, FCI) calculation. 
The single-particle density matrix is defined as
\begin{equation}
 \rho_{pq}^\sigma(t)= \langle \Psi(t) | \hat{c}_{p\sigma}^\dagger \hat{c}_{q\sigma} | \Psi(t) \rangle \;.
 \label{eq:sp-density-matrix}
\end{equation}
For spin-free Hamiltonians, the spatial density matrix is constructed by the spin summation,
\begin{equation}
 \rho_{pq}(t)= \sum_{\sigma}  \rho_{pq}^\sigma(t) = 
 \rho_{pq}^\alpha(t)+\rho_{pq}^\beta(t)\;.
 \label{eq:spin-free-density}
\end{equation}
The natural orbitals are then obtained by a diagonalization of 
the matrix $\boldsymbol{\rho}$ formed by the elements $ \rho_{pq}$ obtained
under time-independent field-free conditions,
\begin{equation}
 \boldsymbol{\rho} \varphi_i^\textup{n} (x)= \nu_i \varphi_i^\textup{n} (x)\;,
 \label{eq:natorb}
\end{equation}
where $\nu_i$ are the natural occupation numbers of the spatial orbitals ($\nu_i \leq 2$, for HF  $\nu_i=[2,0]$),
and $\varphi_i^\textup{n}(x)$ are the corresponding natural orbitals.
It is well-known for electronic ground-state calculations
that CI expansions have favorable convergence properties using the basis set formed
by natural orbitals~\cite{lowdin1955,szabo}.

\begin{figure}
  \includegraphics[width=8.5cm]{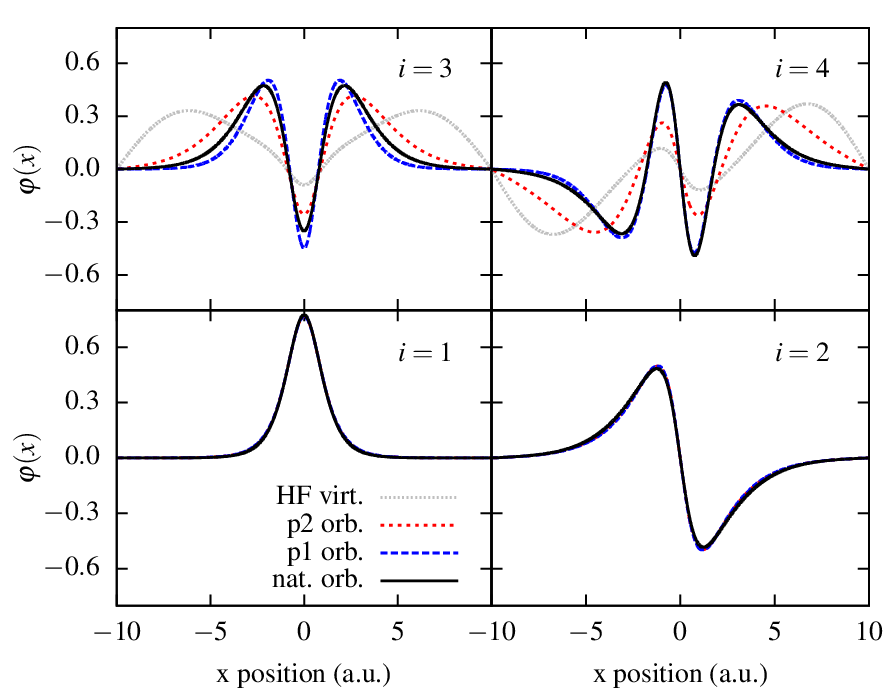}
  \caption{(color online). Single-particle orbitals in the central  region $[-x_c,x_c]$ 
  for 1D beryllium ($N_{\textup{el}}=4$, $x_c=10$).
  Comparison of pseudo orbitals $\varphi_i^{p_1}$, $\varphi_i^{p_2}$ 
  [cf. Eqs.~\eqref{eq:interaction-free-EVP} and \eqref{eq:hn2el}] and natural orbitals $\varphi_i^\textup{n}$
  [Eq.~\eqref{eq:natorb}]. The HF virtuals are additionally plotted in gray for comparison.
  The parameters are the same as in Fig.~\ref{fig:mixed-basis}.
  }
  \label{fig:1dbe-orbitals}
\end{figure}

The four lowest-lying orbitals ($i=1\dots4$) of $\varphi_i^{\textup{p1}}(x)$, $\varphi_i^{\textup{p2}}(x)$
and $\varphi_i^\textup{n}(x)$ for a 1D beryllium model (see Sec.~\ref{ssec:1dbe}) are plotted in 
Fig.~\ref{fig:1dbe-orbitals} together with the virtual orbitals of the HF method (gray, $i=3,4$).
For the occupied orbitals ($i=1,2$, lower panels), there exists, by construction, no difference between the HF and the pseudo orbitals. Only the natural orbitals show a slight modification.
For the convergence of TD-GASCI calculations, however, the virtual orbitals (upper panels, $i=3,4$) are important 
because  of their strong influence on the construction of excited determinants in the
CI expansion.

Whereas the HF virtual orbitals are strongly delocalized, all other types of orbitals remain localized close to the nucleus.
As expected, the highest localization is achieved for the hydrogen-like orbitals $\varphi_i^{\textup{p1}}(x)$ (dashed, blue lines).
The improved pseudo orbitals $\varphi_i^{\textup{p2}}(x)$ show a stronger delocalization (dotted, red lines), the natural
orbitals are in between (black, solid lines). For a discussion of the convergence properties of the TD-GASCI method with respect
to the choice of the orbitals, see Sec.~\ref{ssec:1dbe}. All these orbitals describe a rotated basis for the GASCI expansion and are equivalent 
regarding  completeness (with respect to the underlying FE-DVR basis set).
Thus, if results are converged with respect to
the e-e correlation, the actual choice of these orbitals is not important. 
The choice influences, however, the size of the GAS expansion needed for convergence,
 and therefore for challenging calculations
the accuracy of the simulation.

\subsection{Observables}
\label{ssec:observables}
In the following, we demonstrate the extraction of several observables of relevance for ionization studies from the GASCI
wave function.

The simplest way to extract (single-particle) observables
such as densities in real or momentum space from the GASCI wave function is 
to construct 
the single-particle density matrix $\boldsymbol{\rho}$, Eq.~\eqref{eq:sp-density-matrix} or
Eq.~\eqref{eq:spin-free-density}.
The single-particle spatial density is given by 
\begin{equation}
 n(\boldsymbol{r},t)=\sum_{p,q} \rho_{pq}(t) \varphi_p^*(\boldsymbol{r})\varphi_q(\boldsymbol{r})
 \label{eq:sp-density}
\end{equation}
which transforms to
\begin{equation}
 n(x,t)=\sum_{p,q}\rho_{pq}(t) \varphi_p^*(x) \varphi_q(x)
 \label{eq:sp-density-1d}
\end{equation}
for the case of the 1D partially-rotated basis set.

The momentum distribution of one particle can similarly be computed
by~\cite{hochstuhl2014}
\begin{equation}
 n(\boldsymbol{k},t)=\sum_{p,q} \rho_{pq}(t) \tilde{\varphi}^*_p(\boldsymbol{k}) \tilde{\varphi}_q(\boldsymbol{k})\;,
 \label{eq:sp-momentum}
\end{equation}
with the Fourier transform of the basis functions $\tilde{\varphi}_p(\boldsymbol{k})$.

For the 1D FE-DVR basis functions, Eqs.~\eqref{eq:fedvr-bridge} and \eqref{eq:fedvr-element},
the transformed functions are given for the bridge functions by
\begin{equation}
 \tilde{\varphi}^i_m(k)=\frac{\sqrt{w^i_{n_g-1}+w_0^{i+1}}}{\sqrt{2 \pi}}\exp\left(-ikx_{n_g-1}^i\right)\; ,
\end{equation}
and by
\begin{equation}
 \tilde{\varphi}^i_m(k)=\sqrt{\frac{w^i_m}{2\pi}}\exp\left(-ikx^i_m\right) 
\end{equation}
for the element functions.

Since we are interested in the momentum or energy distributions of photoelectrons,
it is necessary to remove the influence of the potentials of the nuclei.
To this end, we assume a large separation of the electronic wave packet  from
its binding potential and include only functions outside a certain radius $r_\textup{ion}$ into
the calculation of Eq.~\eqref{eq:sp-momentum}. This corresponds to the projection onto plane
waves ignoring the central region. This method is asymptotically 
exact~\cite{madsen2007} and  applicable since we deal only with
single continua in our CAS$^*$ schemes.
Double continua drastically increase the complexity of the problem~\cite{argenti2013}.
Further, the momentum representation of the transformed orbitals for $|x|<x_c$ does not need
to be calculated because typically $r_{\textup{ion}} \gg x_c$.

While the total ionization probability can be obtained by integration of the 
photoelectron spectrum, it is often practical to obtain 
this quantity by the usage of a complex absorbing potential
added to the total Hamiltonian,
\begin{equation}
 H^{\textup{CAP}}(t)=H(t)-iV_{\textup{CAP}} \;.
\end{equation}
Throughout, we use a CAP of the form~\cite{miyagi2013}
\begin{equation}
 V_{\textup{CAP}}(x)=1-\cos\left(\frac{\pi( |x|-r_{\textup{CAP}})}{2(x_s-r_\textup{CAP})}\right)
 \label{eq:v-cap}
\end{equation}
for $|x|>r_{\textup{CAP}}$ with $r_{\textup{CAP}}$ the distance from the 
simulation grid center at which the CAP is  turned on.
The normalization $\mathcal{N}(t)=\langle\Psi(t)|\Psi(t)\rangle$ of the wave function as function of time 
provides then a measure of the total ionization probability~\cite{kulander1987}.
For sufficiently long propagation 
times $t_f$ after the end of the pulse,  the continuum  part of the wave function
has passed $r_\textup{CAP}$  and been absorbed and the total ionization yield is given by   
\begin{equation} 
\mathcal{P}(t_f)=1-\mathcal{N}(t_f) \;. 
\label{eq:ionization-prob}
\end{equation}
Of course, using such an approach, it is not possible to discriminate between different ionization
channels or single, double or multiple ionization.

We mention in passing, that high-order harmonic-generation spectra can be
conveniently obtained from the dipole momentum in the acceleration form~\cite{miyagi2013} and the matrix elements of relevance are calculated in analogy 
to the single-particle potential energy.

\subsection{Summary of simulation method}
In total, the TD-GASCI scheme works as follows
\begin{enumerate}
 \item set up FE-DVR basis (weights $w_i$ and points $x_i$)
       and matrix elements for $\hat{t}$, $\hat{v}$ and $\hat{w}$ for $x \in [-x_{c},x_c]$
 \item construct (pseudo) orbitals in $[-x_c,x_c]$ by HF calculations or CI ground state calculations for the case of natural orbitals
 \item construct FE-DVR basis and matrix elements for $\hat{t}$, $\hat{v}$ and $\hat{w}$ for $x \in [-x_s,x_s]$
 \item rotate the parts of $h_{pq}$ that belong to the central region and 
 parts of $w_{pqrs}$, see appendix~\ref{sec:app-electron-integrals}
 \item construct GASCI initial state for $x \in [-x_s,x_s]$ by ITP
 \item perform TD-GASCI calculation in real time
 \item construct single-particle density matrix $\boldsymbol{\rho}(t)$ and extract observables
\end{enumerate}

\section{Numerical examples}
\label{sec:numerical}
To test and validate the TD-GASCI approach for photo-excitation and ionization phenomena of few-electron atoms, 
we follow a long tradition in time-dependent calculations and  apply the theory to 1D 
models of  atoms~\cite{haan1994,bauch2010,pindzola1991,balzer2010,balzer2010b,balzer2010c,ruggenthaler2009,miyagi2013,miyagi2014,miyagi2014b,sato2013}. 
This allows us to study the convergence properties
in direct comparison with accurate simulations of the TDSE. 

In our model, the Coulomb binding potential of the nucleus is given by the regularized potential
\begin{equation}
 V(x_i)=-\frac{Z}{\sqrt{{x_i^2+s^2}}} \;.
  \label{eq:1d-potential}
\end{equation}
The interaction between two electrons at positions $x_i$ and $x_j$ is analogously given by
\begin{equation}
 V(x_i,x_j)=\frac{1}{\sqrt{(x_i-x_j)^2+s^2}} \;.
\end{equation}
For all situations considered in this work, we use a softening parameter of $s=1$.
Further, we describe the interaction with the external field in the dipole approximation and 
use the length gauge, 
cf. Eqs.~\eqref{eq:td-hamil} and \eqref{eq:sp-hamil}, 
either with a Gaussian half-cycle pulse
 \begin{equation}
 F(t)=F_0 \exp\left[ -\frac{(t-t_0)^2}{2\sigma^2}\right] \;,
 \label{eq:halfcycle}
\end{equation}
or
with an electric field 
with a Gaussian envelope
\begin{equation}
 F(t)=F_0 \exp\left[ -\frac{(t-t_0)^2}{2\sigma^2}\right] \cos [\omega(t-t_0)+\varphi_{\textup{CEP}}]\;.
 \label{eq:efield}
\end{equation}
The maximum amplitude is denoted by $F_0$, the pulse duration by $\sigma$, the photon frequency
by $\omega$, 
and the carrier-envelope phase (CEP) by $\varphi_{\textup{CEP}}$. 

\subsection{2-electron model atom (helium like)}
Let us start with $N_{\textup{el}}=2, Z=2$ which results in a helium-like model system for
which the TDSE is exactly solvable without further approximations. The 
exact results are compared with the results of the  TD-GASCI approach.
We solve the two-particle TDSE by discretizing the two-electron coordinates
$x_1$ and $x_2$ in the same FE-DVR basis set as for the TD-GASCI 
using product states (in analogy to Ref.~\cite{schneider2006}) 
to exclude any influence from a difference in basis sets.
For these brute-force TDSE simulations, no partial rotation of the basis is employed. For the TD-GASCI, we perform the rotation.

\subsubsection{Ground-state}
\label{sssec:1dhe-ge}
\begin{figure}
  \includegraphics[width=7cm]{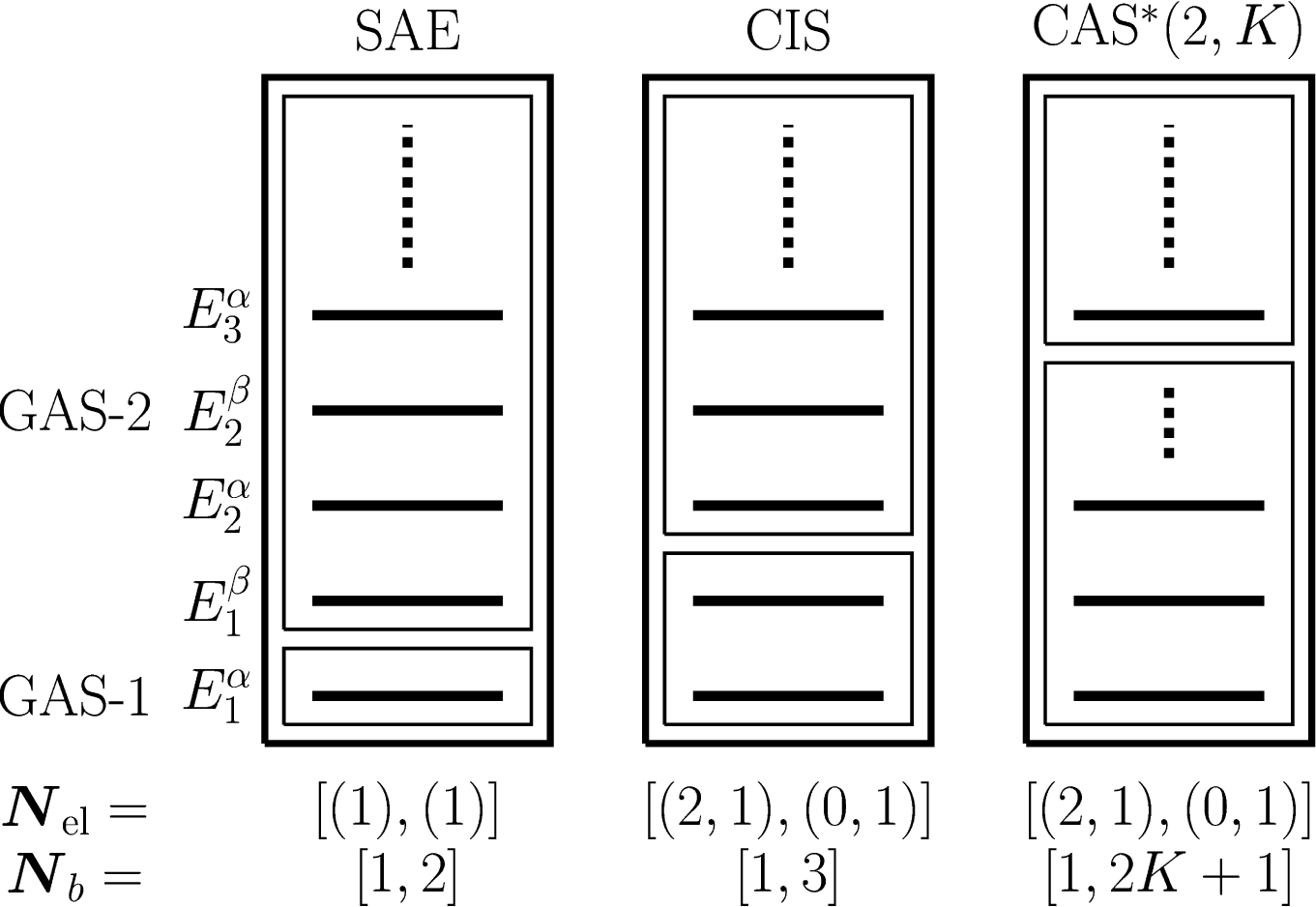}
  \caption{GAS partitions for the two-electron model.
  The acronyms of the different approximations are: ``SAE---single-active electron'', 
  ``CIS---configuration-interaction singles'', ``CAS*(2,$K$)---complete active space'' 
  with a CAS including $K$ spatial orbitals and single excitations outside. 
  See Fig.~\ref{fig:gas-schematic} and Sec.~\ref{ssec:gas-scheme} for notations.
  }
  \label{fig:1dhe-gas}
\end{figure}

The (small) simulation box ranges from $x_s=\pm 15$ 
with a rotated basis to described the central region within $x_c=\pm 10$ for the GAS case. 
The total interval is discretized in $N_e=30$ elements each of which has $n_g=8$ DVR functions. This gives a total of $209$ FE-DVR functions of which $139$ are rotated to pseudo (or natural) orbitals.
The relevant GAS partitions for this two-electron system are sketched in Fig.~\ref{fig:1dhe-gas} together with a description
of the nomenclature, see also Sec.~\ref{ssec:gas-scheme}.

The ground-state energies (GSE) for different GASCI approximations obtained by 
ITP are summarized in Table~\ref{tab:1dhe-ground}.
As expected, the HF approximation GSE is larger than the exact TDSE value. 
We further note that the HF results are
mostly converged with respect to the central region (``center'' vs. ``all'') and only the last 
digit differs. 
By applying the simplest GAS approximations (SAE and CIS), we retain the well-known Brillouin theorem by recovering the HF energy of the 
whole simulation range $[-x_s,x_s]$ exactly.

Adding more pseudo orbitals to the lowest GAS, resulting in a CAS with double excitations up to including $2K$ spin 
orbitals and single excitations above this level [CAS$^*(2,K)$], lowers the GSE. Convergence is achieved for the case 
CAS$^*(2,27)$ with 10557 configurations
in the expansion. This value for the GSE is limited by the choice of $x_c$. By including also the non-rotated part for double
excitations, we recover the TDSE limit exactly up to machine precision (FCI) with 
$43681$ configurations.
We note that with about 10 times less configurations an excellent approximation for the GSE is achieved.

\begin{table}
  \begin{tabular}{lcccll}
    \hline
    Method        &$\boldsymbol{N}_{\textup{el}}$&$\boldsymbol{N}_b$ &  $n_{\textup{conf}}$ & Energy [a.u.] \\
     \hline
     HF center     & -               & -               & $1$     & $-2.22420954$\\
     HF all       & -		    & -               & $1$     & $-2.22420955$\\  
     SAE          &$[(1),(1)]$      & $[1,2]$    & $209$   & $-2.22420955$\\
     CIS          &$[(2,1),(0,1)]$  & $[1,3]$    & $417$   & $-2.22420955$\\
     CAS$^*(2,2)$ &$[(2,1),(0,1)]$  & $[1,5]$    & $832$   & $-2.23617624$\\
     CAS$^*(2,3)$ &$[(2,1),(0,1)]$  & $[1,7]$    & $1245$  & $-2.23747755$\\
     CAS$^*(2,11)$&$[(2,1),(0,1)]$  & $[1,23]$   & $4477$  & $-2.23820292$\\
     CAS$^*(2,27)$&$[(2,1),(0,1)]$  & $[1,55]$   & $10557$ & $-2.23825772$\\
     FCI          &$[(2)]$          & $[1]$      & $43681$ & $-2.23825782$\\
     TDSE         & -               & -               & -       & $-2.23825782$\\
     \hline
  \end{tabular}
 \caption{Ground-state energy as function of GAS for the 2-electron 
 helium-like model.
 ``Center'' refers to a HF calculation for $|x| \leq x_c$ and ``all'' for $|x| \leq x_s$.
  The total number of spin orbitals  is $2N_b=2\times 209$.
 See Fig.~\ref{fig:1dhe-gas} and Sec.~\ref{ssec:gas-scheme} for a definition of the GAS
 spaces and the notations.}
 \label{tab:1dhe-ground}
\end{table}

\subsubsection{Ionization yields and photoelectron spectra}
\label{ssec:1dhe-ionization}
As pointed out in Ref.~\cite{hochstuhl2012}, the TD-RASCI approach 
 allows for an accurate calculation of 
photoionization cross sections including the relevant multiple-excited
states. A systematic investigation of the influence of the 
partially-rotated basis was, however, not carried out.
To test the method against TDSE simulations,
we prepare the 1D helium-like model in its ground-state and shine a 
long Gaussian shaped pulse [Eq.~\eqref{eq:efield}] 
of length $\sigma=100$ and strength $F_0=0.001$ centered at time $t_0=400$ and with $\varphi_{\textup{CEP}}=0$. 
Note that the electric field strength of the rather long pulse is clearly in the 
perturbative regime to avoid saturation of the ionization yield also in the case of
resonant excitation. 
We propagate to a final time of $t_f=4000$ to allow for a reasonable decay 
of all excited resonances. 

To facilitate a large number of calculations 
for different photon frequencies, we choose a rather small system size of $x_s=\pm 40$ 
with the atom centered at $x=0$.
The central region is connected at $x_c=\pm 10$ and a total FE-DVR basis set of $40$ elements with $7$ DVR functions
has been used.
The total ionization yield $\mathcal{P}(t_f,\omega)$ is extracted 
from Eq.~\eqref{eq:ionization-prob} with a
CAP starting at $r_{\textup{CAP}}=20$ in Eq.~\eqref{eq:v-cap}.

In addition to the photoionization with a rather long pulse, we, in a different calculation,
excite the system with a $\delta$-like [$\sigma=0.1, t_0=1, F_0=0.001$ in Eq.~\eqref{eq:halfcycle}]
dipole kick and record the dipole response $\langle x\rangle (t)$ over a long time ($t_f=3000$).
A Fourier transform with respect to the time, 
\begin{equation}
S(\omega)=|\mathcal{F}\left\lbrace x(t)\right \rbrace|^2,
\label{eq:spectrum}
\end{equation}
gives the 
dipole excitation spectrum~\cite{ruggenthaler2009}. 
For better visibility of the positions of the resonances, 
we apply a Blackman window~\cite{blackman-function} to the data before applying the discrete Fourier transform.

\begin{figure*}[t]
 \includegraphics[width=18cm]{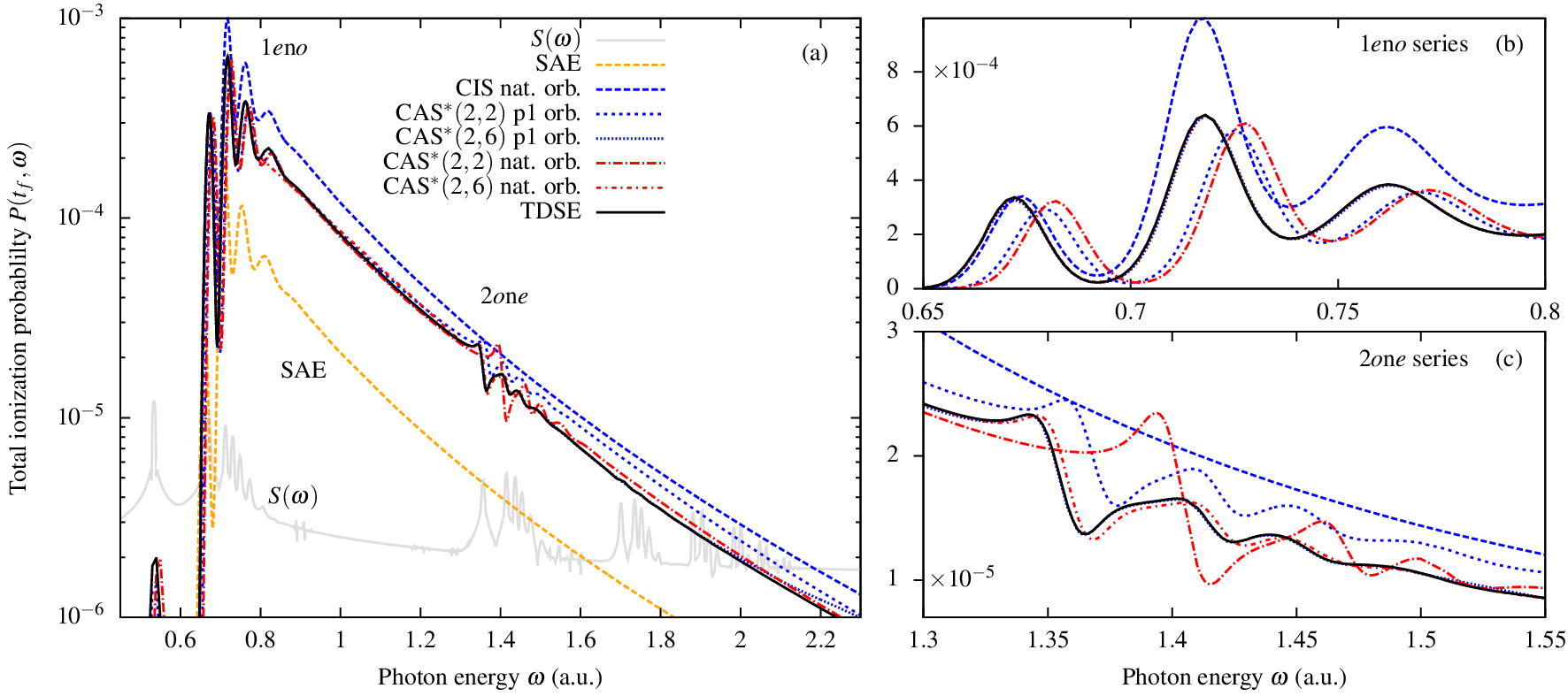}
 \caption{(color online). Ionization probability $\mathcal{P}(t_f=4000,\omega$), cf. Eq.~\eqref{eq:ionization-prob}, of the helium-like
 model for different   GAS approximations as a 
 function of the  photon energy for a fixed pulse duration.
 The results for pseudo orbitals $\varphi^{p_1}$, Eq.~\eqref{eq:interaction-free-EVP}
 and natural orbitals $\varphi^{n}$, Eq.~\eqref{eq:natorb}, in the rotated basis are compared.
  The left panel (a) shows the whole range
 of frequencies on a logarithmic scale and the right panels magnifications of the one-electron excitations (1$e$n$o$, b)
 and the first two-electron resonances (2$o$n$e$, c) on a linear scale. 
 The dipole-excitation spectrum $S(\omega)$, cf. Eq.~\eqref{eq:spectrum},
 for an infinitesimally short pulse from a fully-correlated TDSE simulation 
 is drawn in gray to help identifying the positions of the excited states. The field parameters are $F_0=0.001, \sigma=100, t_0=400$, 
 $\varphi_{\textup{CEP}}=0$ in Eq.~\eqref{eq:efield}.}
 \label{fig:1dhe-ionization}
 \end{figure*}

The ionization yields $\mathcal{P}(t_f,\omega)$ as a function of the photon energy $\omega$ for different GAS approximations 
and the corresponding TDSE result are shown in Fig.~\ref{fig:1dhe-ionization} 
together with the dipole spectrum $S(\omega)$ from 
a TDSE calculation (gray line).
The resulting peaks in Fig.~\ref{fig:1dhe-ionization} (a) can be classified into two groups: (i) single excitations up to $\omega\approx 0.9$
and (ii) double excitations above $\omega \approx 1.2$.
(i) correspond to the excitations of one electron into higher states, where the other electron is still bound
in its ground-state orbital. These are labeled by 1$e$n$o$ where
$e$ ($o$) denotes an orbital that is even (odd) under the parity operation. This series converges to the first ionization threshold $I_p^{(1)}$ for $n\rightarrow
\infty$ and is visible in all GAS approximations, ranging from SAE to the fully converged TDSE result, at approximately
the correct position. We note, however, that the SAE approximation (lower dashed, orange line labeled 'SAE')
underestimates the yield by about a factor of $2$ whereas CIS overestimates the yield (dashed, blue line).

Figure \ref{fig:1dhe-ionization}(b) shows a magnification of the region 
relevant for single excitations (1$e$n$o$) and compares 
the results obtained using different types of orbitals in the central region.
We find that pseudo orbitals $\varphi^{p1}(x)$ (dashed, blue lines) and natural orbitals $\varphi^{n}(x)$ 
(dashed dotted, red lines) describe the single-excitations
well and perfect agreement with the TDSE (black solid line) is achieved for 
CAS$^*(2,6)$, where the GAS consists of an active space of 6 spatial orbitals 
and single excitations above, and practically no difference is visible. 
Further, for the smaller CAS$^*$(2,2) calculation 
with non-converged e-e correlation contributions, the differences 
between pseudo and natural orbitals are only marginal (dotted, blue vs. dashed-dotted, red  lines).

A slightly different picture arises for the two-electron resonances (ii). A zoom-in of 
the 2$o$n$e$ series, i.e., the
simultaneous excitation of one electron into the first excited state and of the other electron to all possible higher
states, is shown in Fig.~\ref{fig:1dhe-ionization}(c).
These resonances are absent for the SAE and CIS approximations and appear only if double excitations
are included into the GAS. 
Again, good agreement with the TDSE is achieved for large CAS$^*$(2,6), however it turns out that
there is a difference in the convergence behavior for natural and pseudo orbitals for small
CAS$^*$(2,2), i.e., not fully correlated calculations. Where the natural orbitals $\varphi_i^n(x)$ (dashed-dotted, red lines)
have problems in describing the correct energy position of the resonances, the pseudo orbitals $\varphi_i^{p1}(x)$
overestimate the overall ionization yield (dotted, blue lines) but predict better excitation energies.

\begin{figure*}
 \includegraphics[width=18cm]{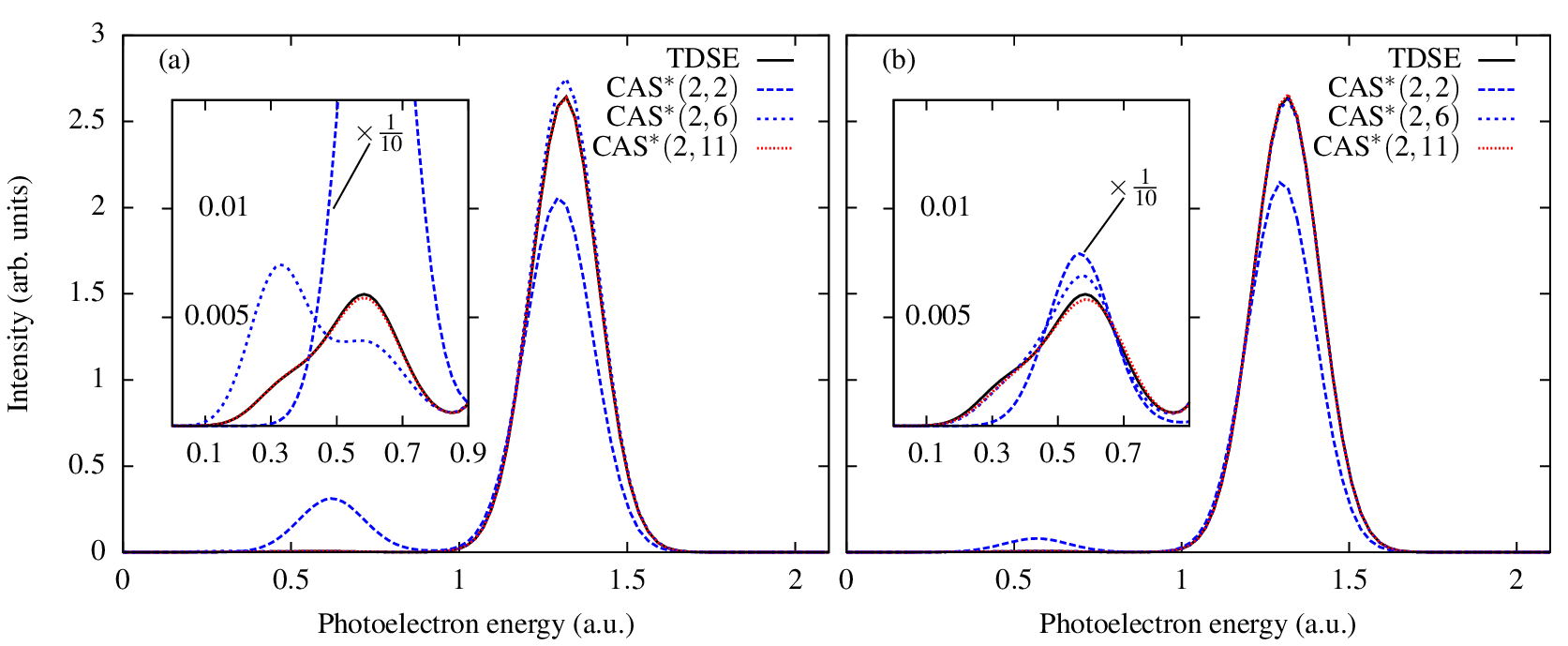}
 \caption{(color online). Photoelectron spectra of the 1D helium-like model for a 
 short pulse [Eq.~\eqref{eq:efield}] with $\sigma=5$, $F_0=0.01$, $\varphi_{\textup{CEP}}=0$
 and a photon energy of $\omega=2.1$ using (a) pseudo orbitals $\varphi_i^{p_1}(x)$ and
 (b)  natural orbitals $\varphi_i^{n}(x)$. The CAS$^*(2,2)$ results are
  scaled by a factor of 1/10 in the inserts.}
 \label{fig:photoionization-1dhe}
\end{figure*}

This behavior is even more pronounced for calculations of the photoelectron spectra, which are 
shown in Fig.~\ref{fig:photoionization-1dhe}. 
The spectra were obtained with the method described in Sec.~\ref{ssec:observables} 
(see also Ref.~\cite{hochstuhl2014}), and a radius of $r_{\textup{ion}}=20$ was used 
for ionization.
Calculations were performed for  $F_0=0.01$,
 $\sigma=5$, $\varphi_{\textup{CEP}}=0$  and $\omega=2.1$ in Eq.~\eqref{eq:efield}, which results in a rather broad excitation bandwidth.
The results for pseudo orbitals are shown in Fig.~\ref{fig:photoionization-1dhe}(a) and for natural orbitals in  Fig.~\ref{fig:photoionization-1dhe}(b),
together with the TDSE result (black line).
The insets show a magnification of the correlation satellites (``shake-up'')
at lower photoelectron energy which are nearly invisible in the total spectra.
In these processes, the photon energy is shared between the photoelectron and
a second, still bound electron. The result is a slower photoelectron, which gives the 
correlation peak in the energy distribution, and an ion in an excited state.

The main peak at an energy of $1.3$ is well-described in position and shape by both types of orbitals and the different 
CAS$^*$ approximations considered. For small active spaces as in CAS$^*$(2,2) 
(dashed, blue line), 
this peak is drastically underestimated for both types of orbitals. An even more pronounced
influence of the CI truncation can be observed in the satellites below an energy of about $0.8$ (inserts).
Note that the case of CAS$^*(2,2)$ has been scaled by a factor of $0.1$ in the 
inserts.

For a limited active space the choice of the orbitals becomes vital and natural orbitals
describe the shape of the peak and its magnitude better. Especially the excitations for CAS$^*(2,6)$ 
into higher orbitals (lower resulting photoelectron energy) is significantly closer to the TDSE result than the CAS$^*$(2,2) results.
Since both choices of orbitals represent rotations in the space of virtual 
(i.e.~unoccupied) HF orbitals and both form a complete single-particle basis,
the results converge toward the TDSE solutions
in the limit of a large active space  [red dotted lines for CAS$^*(2,11)$].

\subsection{4-electron model atom (beryllium like)}
\label{ssec:1dbe}
We now consider the more complex model with $N_{\textup{el}}=Z=4$ in 
Eq.~\eqref{eq:1d-potential}, which results in a 
beryllium-like 1D model. It can be solved exactly only for very 
special situations, e.g., with TD-FCI or TDSE simulations for very small simulation boxes 
and single-particle basis sets.

\begin{figure}
  \includegraphics[width=7cm]{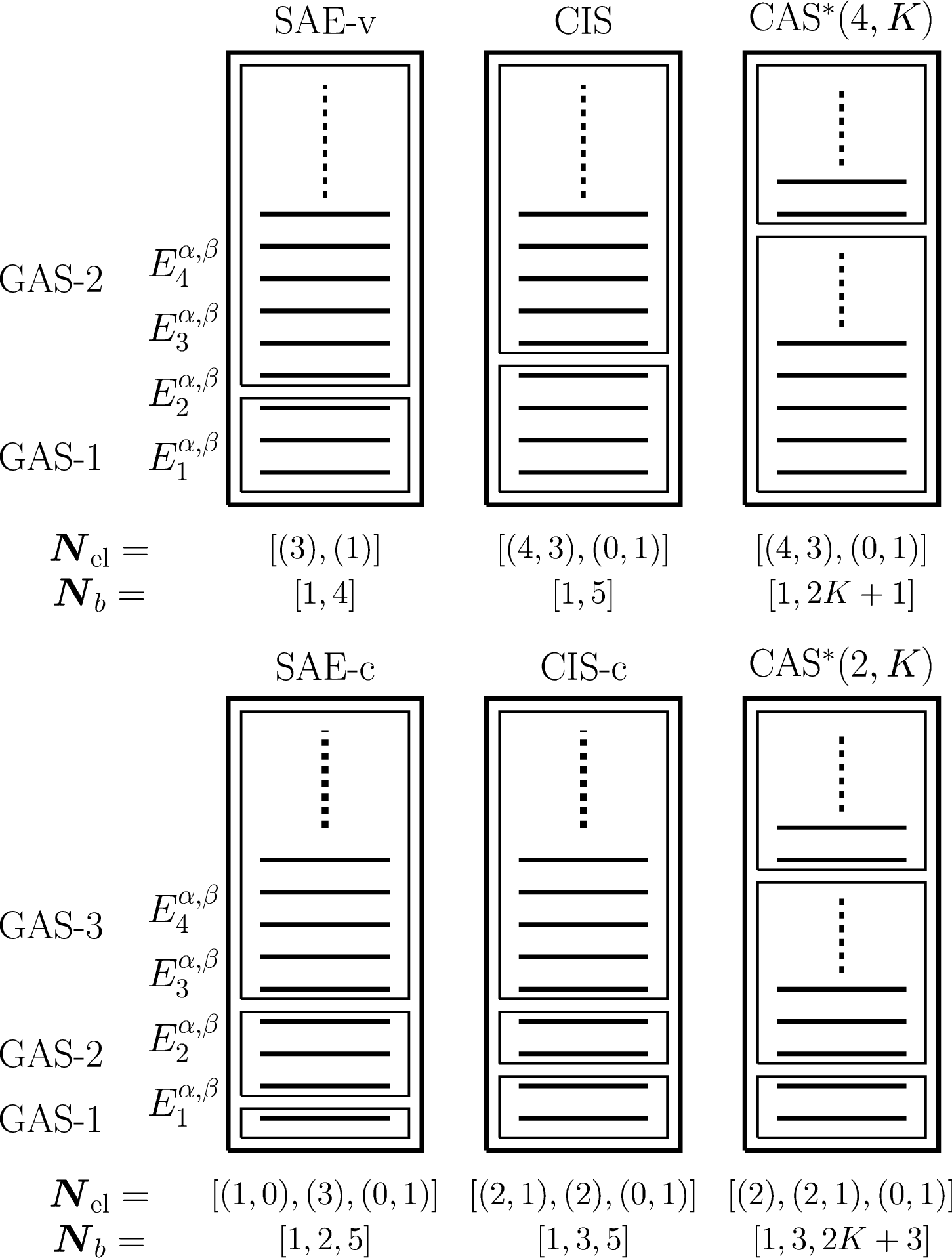}
  \caption{Schematics of the GASs for the four-electron beryllium-like model. 
  The label v (c) refers to an active valence (core) orbital. 
   CAS$^*(2,K)$ and CAS$^*(4,K)$ are active
  spaces with two and four electrons, respectively, with single excitations out of the CAS.
  The case CIS-v equals CAS$^*(2,1)$.
    See also Fig.~\ref{fig:cas-schematic} and Sec.~\ref{ssec:gas-scheme}.}
  \label{fig:1dbe-gas}
\end{figure}
The relevant GAS partitions are shown in Fig.~\ref{fig:1dbe-gas}. In contrast to helium, the 
four electrons occupy the two lowest-lying spatial orbitals, which we will refer to as ``core'' (c)
and ``valence'' (v) orbitals in the following. 
Thus, we can define SAE  approximations for the 
core and the valence orbital, respectively.
In analogy, we can define CIS-like approximations and active spaces with two [CAS$^*(2,K)$] or 
all four [CAS$^*(4,K)$] electrons being active. 
For CAS$^*(2,K)$, the inner-shell electrons are frozen and for the outer-shell electrons, double 
excitations up to spatial orbital $K$ are included. For CAS$^*(4,K)$, analogously, all electrons
can occupy the $K$ spatial orbitals, which also includes 4-fold excitations. For both situations,
single excitations out of the CAS are included.

\subsubsection{Ground-state}
\label{sssec:1dbe-ge}
The GSEs as a function of the GAS partition are collected 
in Table~\ref{tab:1dbe-ground} for different pseudo orbitals $\varphi_i^{p1}(x)$ and
$\varphi_i^{p2}(x)$, cf. Eqs.~\eqref{eq:interaction-free-EVP}-\eqref{eq:hn2el}, and 
a multi-configuration time-dependent Hartree-Fock (MCTDHF) calculation~\cite{hochstuhl2010}.
The parameters for the simulation box ($x_s$ and $x_c$) and the FE-DVR basis
are the same as for helium; Sec. IV.A.

\begin{table}
 \begin{tabular}{llcc}
   \hline
   Approx.      &  $n_{\textup{conf}}$ & $E_0^{\textup{1}}$ & $E_0^{\textup{2}}$ \\
   \hline
   HF $(-x_c,x_c)$    & $1$       & \multicolumn{2}{c}{$-6.73941916$} \\
   SAE-v        & $208$    & $-6.73943439$ & $-6.73943439$ \\
   SAE-c        & $208$    & $-6.73941916$ & $-6.73941916$\\
   CIS-v        & $415$    & $-6.73944960$ & $-6.73944960$\\
   CIS-c        & $415$    & $-6.73941916$ & $-6.73941916$\\
   CIS          & $829$    & $-6.73944961$ & $-6.73944961$\\
   CAS$^*(2,2)$ & $828 $   & $-6.77002039$ & $-6.76960858$\\  
   CAS$^*(2,3)$ & $1239$   & $-6.77375320$ & $-6.77266039$\\ 
   CAS$^*(2,21)$& $8295$   & $-6.77486786$ & $-6.77486757$\\ 
   CAS$^*(2,41)$& $15375$  & $-6.77486825$ & $-6.77486825$\\ 
   CAS$^*(4,3)$ & $3717$   & $-6.77793224$ & $-6.77428136$\\ 
   CAS$^*(4,4)$	& $9876$   & $-6.78325375$ & $-6.77940715$\\ 
   CAS$^*(4,10)$& $181125$ & $-6.78491205$ & $-6.78439562$\\ 
   MCTDHF~\cite{hochstuhl2010}         &   $10^\textup{a}$     & \multicolumn{2}{c}{ $-6.7851$} \\
 \hline
 \end{tabular}
 \caption{The same as Table~\ref{tab:1dhe-ground} but for beryllium. The energies $E_0^1$ and $E_0^2$
 correspond  to the pseudo orbitals $\varphi_i^{p1}$ and $\varphi_i^{p2}$, respectively.
 \newline \footnotesize{$^\textup{a}$In this method, the orbitals and thus the configurations are time dependent.} }
 \label{tab:1dbe-ground}
\end{table}

As expected, the HF GSE is above the fully correlated reference result. 
The two SAE approximations give an impression of the influence of the choice of $x_c$.
Where an active core orbital (SAE-c) gives exactly the same GSE as the HF result up to 
numerical precision (which is a 
manifestation of the Brillouin theorem),
an active valence orbital (SAE-v) lowers the GSE. This can be understood
by the larger spatial extension of the valence orbital in comparison to the core orbital.
The former exceeds the central region, for which the HF calculation was performed while
the strongly localized core orbital is completely captured within the region $\pm x_c$.
During the ITP of the TD-GASCI equations, the initial wave function constructed from 
the valence orbital is allowed to relax on the 
increased grid. This results in a lower GSE, even if only single excitations are included.
The error in the GSE due to the choice of $x_c$ is on the order of $10^{-5}$ for
these parameters. 
A similar observation can be made for the CIS-v and -c  approximations with an active
valence or core orbital. 
The lowest energy for CIS is obtained when all four electrons are allowed to relax 
on the  entire simulation grid.

For the GAS partitions with only single excitations, the choice of the virtual space,
i.e., the rotated orbitals within $(-x_c,x_c)$ is unimportant because 
all orbitals are included on the same level. Therefore, both types of pseudo
orbitals give exactly the same value for the GSE.
The account for correlations, either by two or four active electrons, changes this picture.
Two limits of e-e correlations can be defined: (i) with frozen core [CAS$^*(2,K)$]
and (ii) with all electrons active [CAS$^*(4,K)$].
For (i), the lowest energy 
is reached for about $K=41$, where both types of pseudo orbitals
converge to the same result and an increase of the active space does not change the GSE.
For smaller active spaces, however, we observe a better, i.e. lower, ground-state
using the hydrogen-like pseudo orbitals $\varphi^{p_1}(x)$. This effect is seen 
most clearly  for the
first correction to the CIS result, CAS$^*(2,2)$.
For (ii) with four active electrons, the number of configurations increases dramatically due to the
exponential scaling, cf. Eq.~\eqref{eq:exp-conf}, and
the GSE is lowered significantly. Again, better results are obtained with the pseudo orbitals of type
$\varphi^{p_1}(x)$.

Finally, we note that our method with 181125 configurations, CAS$^*(4,10)$, 
does not reach completely 
the fully-correlated GSE of the MCTDHF calculation, where in addition
to the expansion coefficients of the wave function also the single-particle orbitals
are allowed to relax.
In contrast to TD-GASCI, the MCTDHF method considers a FCI approach with time-dependent orbitals.
Thus, for advancing in time, in addition to the expansion coefficients $C_I(t)$, like in TD-GASCI,
also the orbitals need to be propagated. This results in a non-linear, numerically complex and 
demanding scheme of which the properties for time-dependent calculations in the context of 
photoionization remain to be fully explored.
Further, MCTDHF calculations are feasible for $N_{\textup{el}}\lesssim 10$ with highly-optimized
codes.  Currently, progress towards larger systems is made in the combination of restricted-active spaces time-dependent orbitals~\cite{miyagi2013,miyagi2014,miyagi2014b}.

\subsubsection{Excitation spectra}
\label{sssec:beexc}
We now turn our attention to the time-dependent properties of TD-GASCI 
by considering the dipole excitation spectrum $S(\omega)$, cf. Eq.~\eqref{eq:spectrum},
of the 1D 4-electron beryllium-like model. The spectra are calculated
by exciting the system with a small $\delta$-kick
of the ground-state wave function and the Fourier transform of the time-dependent dipole $\langle x (t) \rangle $,
cf. Sec.~\ref{ssec:1dhe-ionization} for method and parameters.

\begin{figure}[t]
 \includegraphics[width=8.5cm]{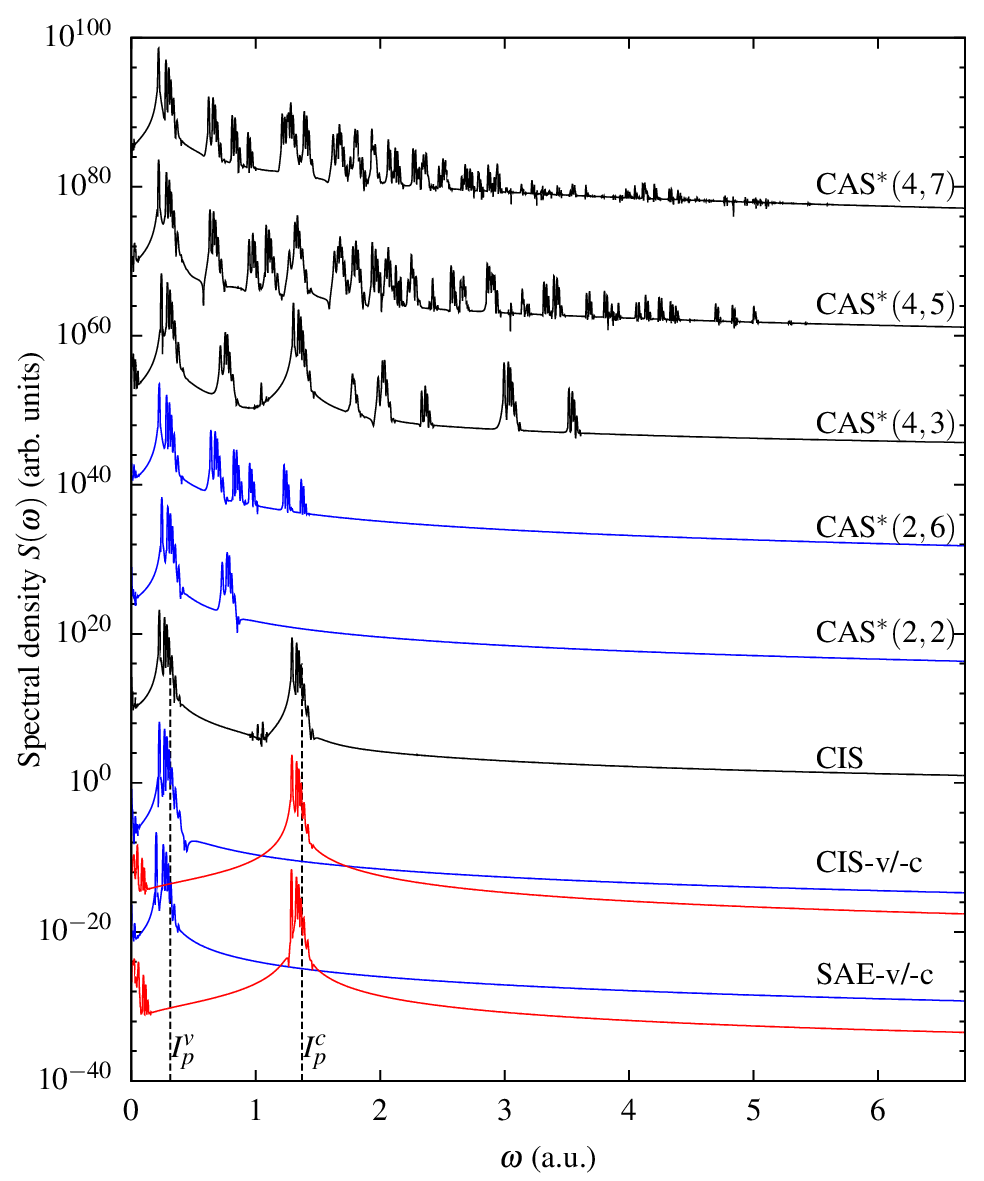}
 \caption{(color online).  Dipole excitation spectrum $S(\omega)$, cf. Eq.~\eqref{eq:spectrum},
  of the 1D 4-electron beryllium-like model for an excitation of $\sigma=0.1$, $t_0=1$, $F_0=0.001$ [Eq.~\eqref{eq:halfcycle}] in different 
 GAS approximations; shown is the total energy range from the ground-state energy to 
 full four-fold ionization ($I_p^4=-E_0$).
 The first ionization potentials for ionization from the valence orbital, $I_p^v$, and core orbital $I_p^c$, are indicated by dashed
 vertical lines. The labels ``v'' and ``c'' refer to the valence and core orbital, respectively.
 Red lines (higher energy, limited by $I_p^c$), correspond to core  electrons only, blue is for the valence shell (lower energy, $I_p^v$).
 }
 \label{fig:dipole-be}
\end{figure}
The results for various GAS partitions are compiled in Fig.~\ref{fig:dipole-be}. 
We define ionization potentials, $I_p^v$ and $I_p^c$, for the valence and the
core orbitals, respectively. In first approximation, they are according to Koopman's theorem
given by the corresponding energy of the occupied HF orbitals.
For the SAE approximations, the ionization potentials
are recovered in the dipole spectrum by a series of excitations,
which converge toward $I_p^{v/c}$ (dashed vertical lines in Fig.~\ref{fig:dipole-be}). 
A similar behavior is found for the CIS approximation of the valence and the core electrons.
For the complete CIS calculation, both series are resolved, i.e., excitation from the core
and the valence orbital is possible. However, not two electrons simultaneously, which results in structureless
continua between $I_p^v$ and $I_p^c$ and above $I_p^c$. 

In these regions, multi-electron resonances appear as a consequence of the allowance for multiple excitations
in the GAS partition. For frozen-core calculations, CAS$^*(2,K)$, additional peaks arise above
$I_p^v$ due to the simultaneous excitation of two valence electrons into a doubly excited state
and its subsequent decay with one electron in the continuum.
The spectra become much more complex, if all four electrons are active, CAS$^*(4,K)$.
For these, doubly, triply and quadruply excited states are accessible and appear as multiple-excited
resonances in the dipole spectrum. These excitations converge towards an energy where all four electrons are liberated 
($I_p^{(4)}=-E_0 \simeq 6.78$).
Thus, besides its computational advantages and systematic
approach to e-e correlation effects,
TD-GASCI allows additionally for a clear interpretation of excitation spectra
in terms of systematic adding of configurations to the expansion~\eqref{eq:gas-expansion}.

\subsubsection{Orbital influence on  TD-GASCI convergence}
\label{sssec:be-orbitals}

\begin{figure*}[t]
 \includegraphics[width=17cm]{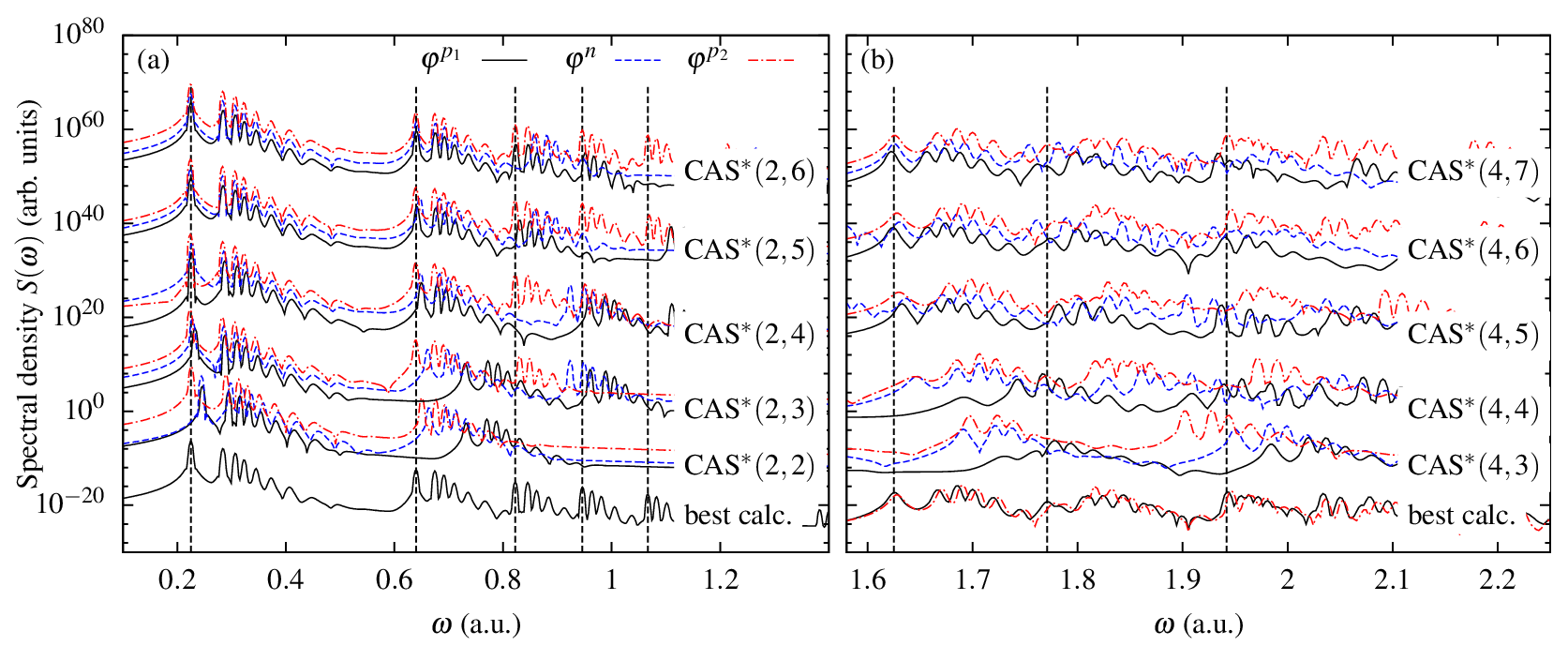}
 \caption{(color online). Parts of the dipole excitation spectrum $S(\omega)$ of the 4-electron 
 beryllium-like model  for different GAS approximations and orbital basis sets in the central region.
 (a) fixed core, (b) all electrons are active. Dashed vertical lines are guides to the eye for reference to the best 
 [CAS$^*$(2,11) for (a) and CAS$^*$(4,7) for (b)] approximation 
 (bottom  line). The individual lines are vertically shifted for better visibility. The full spectrum and parameters are given in Fig.~\ref{fig:dipole-be}.
 }
 \label{fig:1dbe-dip-orb}
\end{figure*}

In Sec.~\ref{sssec:1dbe-ge}, we discussed the influence of the type of the pseudo orbitals
on the GSE of the system and found that $\varphi^{p_1}(x)$ [Eq.~\eqref{eq:interaction-free-EVP}] outperform 
$\varphi^{p_2}(x)$ [Eq.\eqref{eq:hn2el}] for ground-state calculations. 
In Fig.~\ref{fig:1dbe-dip-orb}, the dipole spectra $S(\omega)$
for different CAS$^*$ approximations
and the three types of orbitals, $\varphi^{p_1}(x)$, 
$\varphi^{p_2}(x)$ and $\varphi^{n}(x)$ [Eq.~\eqref{eq:natorb}] are shown.
Figure \ref{fig:1dbe-dip-orb}(a) shows GAS approximations with 
two active electrons in the spectral region below the excitation energy of core electrons, 
cf. Fig.~\ref{fig:dipole-be}, in which the energies of the single- and double excitations 
of the valence electrons are located. The lowest black line shows the converged result 
obtained by a  CAS$^*(2,11)$ calculation and dashed vertical lines
the lower threshold energy of each series as a guide to the eye.

The first series corresponds to the one-electron excitations and is well represented in all CAS$^*$ approximations
for each type of orbitals. We notice, however, that the $\varphi^{p_2}(x)$ pseudo orbitals of type 2 (dashed-dotted, red lines)
have better convergence properties and reproduce the correct position in energy 
already in the lowest CAS$^*$ approximation.
For the two-electron resonances the influence of the orbital choice becomes more pronounced. For all considered
approximations, the pseudo orbitals $\varphi^{p2}(x)$ perform better, and the higher-lying series
are closer to the converged result. A similar statement can be made for natural orbitals with respect to the
first double-excitation series, however higher series are more off the correct result.
The worst result is obtained with the pseudo orbitals of type $\varphi_i^{p1}(x)$, which are only able 
to reproduce resonances at the correct positions if the active space is much larger than that of the other orbitals.

The case of four active electrons above $I_p^c$ is shown in panel (b), where higher excited resonances appear in 
the spectrum. Again, the best result for CAS$^*(4,7)$ is shown in the bottom. Here, due to the complex spectrum, a 
clear classification of the orbitals is difficult. However, we find that also for this case the
improved orbitals $\varphi^{p_2}(x)$ perform well and predict excitations at the correct positions. 
For the calculation of 4-fold excitations, the choice of the orbitals is less important and active spaces chosen too 
small result in wrong excitation energies for all orbitals, also the improved ones.
However, we note that $\varphi^{p_2}(x)$ are
especially designed for double excitations of the valence electrons by considering the $N_{\textup{el}}-2$ electron problem for
the calculation of an effective potential. Generalizations of this scheme to orbitals
calculated from $N_{\textup{el}}-3$ or $N_{\textup{el}}-4$ potentials in order to describe the removal of two or more electrons
in combination with excited states of the ion are difficult. The main problem is that such generalized single-particle 
orbitals need to describe the removal of a single electron accurately in addition to the above-mentioned effects.

In total, the pseudo orbitals of type $\varphi^{p2}(x)$ outperform 
natural $\varphi^{n}(x)$ 
and type $\varphi^{p1}(x)$ 
pseudo orbitals in time-dependent excitation scenarios if two-electron excitations are 
considered. We expect this favorable property of the $\varphi^{p2}$ type orbitals to improve 3D calculations for 
real atoms and molecules as well.

\subsection {Molecular model systems}
\label{sec:molecules}
To demonstrate the generality of the TD-GASCI approach, we present in the following  a study of the  
ground-state energy and the nonperturbative dynamics of a diatomic molecule in a strong field. 

Consider, for each electron, the one-dimensional diatomic potential consisting of two atomic species 
\begin{equation}
 V(x,R)=-\frac{Z_1}{\sqrt{(x-\frac{R}{2})^2+1}}-\frac{Z_2}{\sqrt{(x+\frac{R}{2})^2+1}},
\end{equation}
with $x$ the electron coordinate, $R$ the internuclear distance, and $Z_i$ $(i=1,2)$ the nuclear charges. 
A two-electron hydrogen-like molecule is then defined by $Z_1=Z_2=1$ and a four-electron lithium-hydride equivalent by $Z_1=3$ and $Z_2=1$.
Such models are well-established in the literature, 
e.g., ~\cite{balzer2010,balzer2010b,balzer2010c,tolstikhin2011,sato2013}.
We point out that for $x_c \gg R$, the choice of the grid reference, i.e. center of mass, center of charge or 
the geometric center of the molecule, does not influence the calculations. For the calculation
of absolute values for dipoles, however, this reference has to be taken into account.

\subsubsection{Ground-state properties}
\label{ssec:mol-gs}
The total energy of the system, corresponding to the Born-Oppenheimer energy surface, is calculated by
\begin{equation}
 E_t=E_\textup{el}+\frac{Z_1Z_2}{\sqrt{R^2+1}}\;,
\end{equation}
where $E_\textup{el}$ denotes the total electronic ground-state energy. We note that in contrast to Ref.~\cite{balzer2010} the internuclear repulsion is also regularized. 
This is necessary to treat both interactions on a similar footing
and obtain a correct convergence towards the dissociation limit, $E_d$.

\begin{figure}
 \includegraphics[width=8cm]{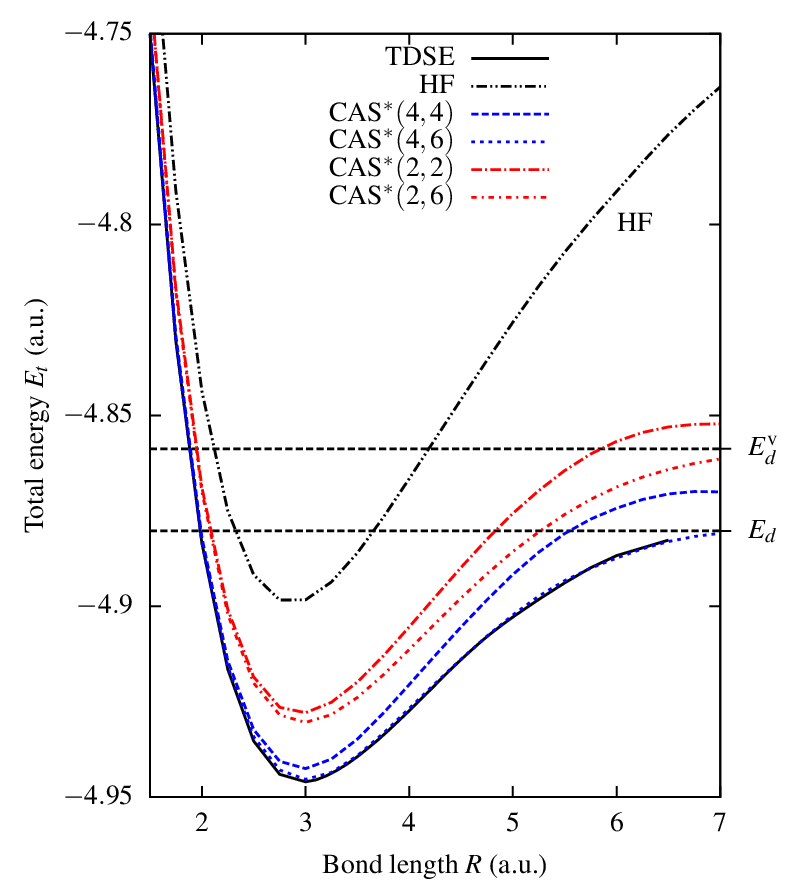}
 \caption{(color online). Total energy of the 1D 4-electron LiH-like model. 
 The dissociation limit is indicated by dashed 
 lines for fully relaxed Li ($E_d$) and for Li with fixed core ($E_d^\textup{v}$).
 See text for details. Dotted-dashed and dotted long-dashed, red lines converging to $E_d^{\textup{v}}$
 are for two active [CAS$^*$(2,.)] electrons, blue lines converging to $E_d$ 
 for all four being active [CAS$^*$(4,.)].
 The TDSE result (full, black) is for a smaller FE-DVR basis set~\cite{balzer2010}.}
 \label{fig:LiH-binding}
\end{figure}

The total energy of the 1D 4-electron LiH-like model as function of the internuclear distance $R$ is shown in Fig.~\ref{fig:LiH-binding}.
Parameters for the calculation are $x_s=\pm 50$ and $x_c=\pm 10$. The box is discretized by 50 elements of which each contains 8 DVR functions.
The GAS nomenclature is as for  the 4-electron atomic model, see Fig.~\ref{fig:1dbe-gas}.
For this prototype four-electron model molecule, reference results are available in the literature~\cite{balzer2010}.
As expected, the closed-shell restricted Hartree-Fock code does not predict the correct dissociation threshold $E_d$ for Li and H in their corresponding ground-state.
Similar behavior is observed for the SAE-v and CIS(-v) approximation in this basis (not shown in the figure).
Including more configurations in the central region, however, repairs this behavior and the potential energy curve converges quickly (for only 4 additional spatial orbitals in the active space) towards the 
four-particle reference TDSE results.

Two different dissociation thresholds, i.e., the ground-state energy of the fragments at 
infinite internuclear distance, are indicated in Fig.~\ref{fig:LiH-binding}:
$E_d$ corresponds to a FCI calculation of Li ($|Z|=N_{\textup{el}}=3$) and $E_d^{\textup{v}}$ to a calculation,
where the $1$s level in Li was fixed and only the unpaired valence electron was
allowed to relax.
For GAS calculations with fixed inner shell electrons and only 2 active, 
CAS$^*(2,K)$, the dissociation limit of the 
LiH molecule corresponds to the energy $E_d^\textup{v}$ and is correctly reproduced by including about 
$6$ spatial orbitals in the active space.

\subsubsection{Strong-field ionization}
\label{ssec:mol-ion}
In this section, we illustrate the potential of the TD-GASCI method by studying the influence 
of electron-electron correlation on the preferred 
direction of electron ejection 
with respect to the external field and molecular orientation
in the heteronuclear polar diatomic LiH-like 1D model molecule subject to strong-field ionization at 800 nm. 
There is currently a strong interest in the elucidation of this question. 
For example in the OCS molecule, experiments with circularly polarized light and theory  
show that   ionization 
occurs most readily from the O-end, i.e., when the field points towards the 
S-end~\cite{holmegaard2010,dimitrovski2011,madsen2013}. For linearly polarized 
light, on the other hand, one experiment reports most ionization from the 
S-end~\cite{ohmura2014}, while another most perpendicular 
to the molecular axis~\cite{hansen2012}.
For the CO molecule, as another example, 
strong-field ionization experiments performed in the tunneling regime
report that ionization occurs most readily when the external field has a component pointing 
from the C- to the O-end, and the electron leaves from the C-end~\cite{ohmura2011,li2011,wu2012}.

This is in contrast with the results from application~\cite{madsen2012,madsen2013} of SAE
approximation tunneling theory~\cite{tolstikhin2011},  which predicts that ionization is most
likely when the field points from the O- to the C-end. Recently, many-electron effects expressed
in terms of dynamic core polarization as accounted for at the TDHF mean-field level of theory were
shown to improve the agreement between experiment and theory~\cite{zhang2013}. Also, in the future, many-electron 
effects may be addressed by application of many-electron tunneling theory~\cite{tolstikhin2014}.  Clearly, the TD-GASCI approach is particularly well-suited for an investigation of  many-electron effects on the preferred electron ejection direction
since e-e correlation can be added in a controllable manner by suitably extending the active space.

\begin{figure}
 \includegraphics[width=8cm]{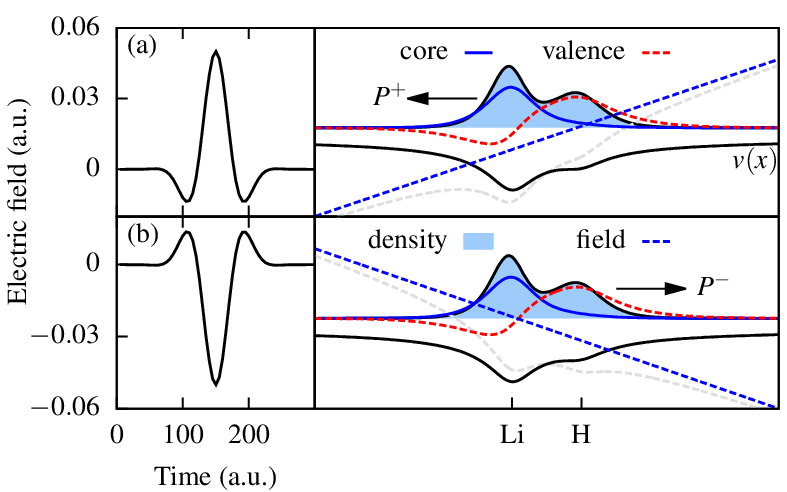}
 \caption{(color online). Sketch of the scenarios considered with the 1D 4-electron LiH-like model 
 molecule under strong-field ionization. 
 The electric field $F(t)$ is shown in the left part for  (a) $\varphi_\text{CEP}=0$ and 
 (b) $\varphi_\text{CEP}=\pi$. The two doubly occupied core and valence
 HF orbitals are plotted along with the single-particle initial density  and potential. 
 For better visibility of the densities, the dipole potential (diagonal dashed lines) 
 at time of maximum electrical field is magnified in the figure.
 }
 \label{fig:ionization-direction}
 \end{figure}

We begin the study by preparing the LiH-model molecule in its electronic ground-state at the equilibrium distance of $R=3$, cf. Sec.~\ref{ssec:mol-gs}.
A short Gaussian-shaped single-cycle [Eq.~\eqref{eq:efield}] 800 nm pulse ($\omega=0.57$)
of duration $\sigma=30$ with electrical field amplitudes 
of (i) $F_0=0.025$ and (ii) $F_0=0.05$ excites the system. For a fixed orientation of the molecule, the peak of the field can be oriented towards the nucleus of either Li or H, depending on on the carrier envelope phase, $\varphi_\text{CEP}$. In Fig.~\ref{fig:ionization-direction} the considered cases 
$\varphi_\text{CEP} =0$ and $\varphi_\text{CEP}=\pi$ are sketched.

\begin{figure}
 \includegraphics[width=8.5cm]{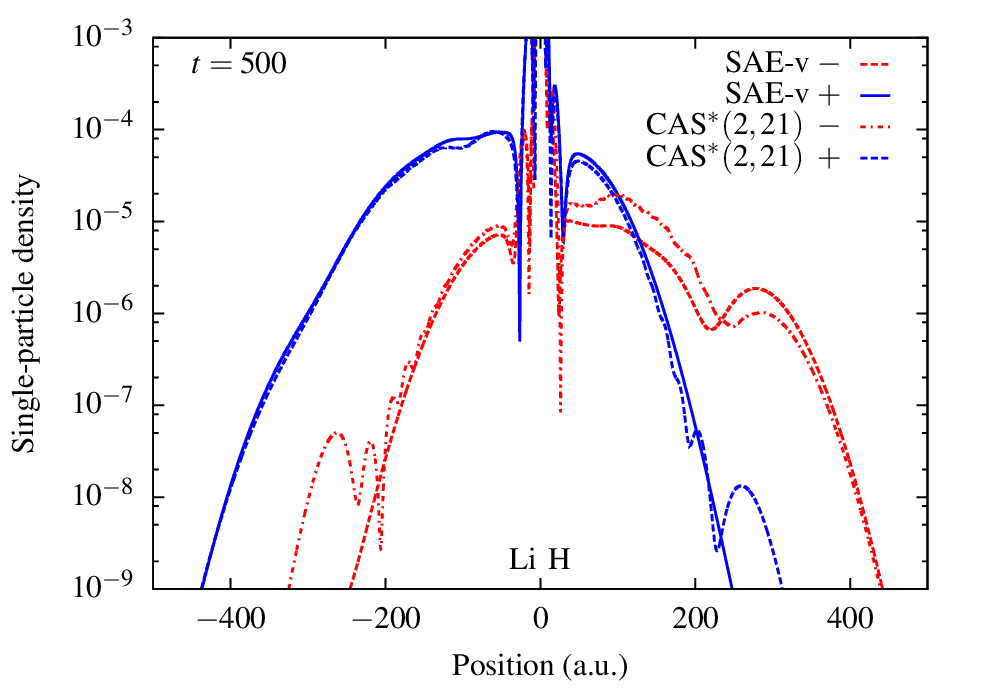}
 \caption{(color online). Single-particle density $n(x)$, cf. Eq.~\eqref{eq:sp-density-1d}, of the 
 1D 4-electron LiH-like molecule exposed to single-cycle pulses (i) after a propagation of $t=500$.}
 \label{fig:lih-ionization-direction}
\end{figure}

To calculate the total ionization yield $\mathcal{P}(t_f)$, cf. Eq.~\eqref{eq:ionization-prob},
for a given $\varphi_\text{CEP}$, a CAP [Eq.~\eqref{eq:v-cap}],  which removes liberated electrons from the simulation box
of size $|x_s|=200$, is placed at a distance of $r_{\textup{CAP}}=100$
from the center of the grid. 
We checked carefully for the influence of the CAP parameters on the 
observable and compared to simulations with very large box sizes without a CAP (cf. Fig.~\ref{fig:lih-ionization-direction})
and no significant change of the results presented were observed. Results are shown for pseudo orbitals of type 1
with $|x_c|=10$. We redid part of the calculations with type 2 orbitals and obtained similar results for the
limit of large CAS$^*$ spaces.

The equations of motion are propagated to a final time of $t_f=15000$ which allows for slow electrons to reach the absorber and thus record the total ionization yields,
${\cal P}^\pm$, for positive ($+$) and negative ($-$) peak electric field amplitudes.
We define  the ratio
\begin{equation}
 \eta=\frac{\cal P^-}{\cal P^+},
 \label{eq:eta}
\end{equation}
which is smaller (larger) than one  if it is more (less) likely to ionize for the situation in the top panels in Fig.~\ref{fig:ionization-direction}, than in the bottom panels.
Furthermore, $\eta=1$ is obtained in the case of equal ionization probability $\cal P^+=\cal P^-$ which is the case for the homonuclear molecules.

The results for $\eta$ for different GAS approximations ranging from SAE to including up to 
21 orbitals in the active space are given in Table \ref{tab:ionizationdirection} for both 
electrical field strengths (i) and (ii).
Let us first discuss LiH at the lower intensity (i). For all approximations, 
ionization is favored when $\varphi_\text{CEP} = 0$ and $F(t)$ is positive at its maximum. In this case  the electron is 
liberated in the direction of Li  [Fig.~\ref{fig:ionization-direction}(a)].
This preference for ionization in this relative geometry  is largest for the simplest possible and most commonly used SAE
approximation. Correlation effects shift 
this result toward more symmetry in the ionization dynamics by a factor of approximately two.
We further note that an active core orbital [CAS$^*(4,K)$, right in Table~\ref{tab:ionizationdirection}]
does not strongly impact the results since the active core and fixed core [CAS$^*(2,K)$, left in Table~\ref{tab:ionizationdirection}] results are quite similar.

By increasing the laser intensity, case (ii), a corresponding behavior is observed, 
but with a much less pronounced favored direction of 
ionization ($\eta$ is larger). This can be explained by the drastically increased 
total ionization yield compared to (i) due to a field strength $|F_0|=0.05$ which is
above the over-the-barrier field strength of about $I_p^2/4 = 0.034$ for
the valence orbital.
For that case, any preference of direction of the electron emission is suppressed.

To learn more about the effect of e-e correlation, we additionally performed calculations on 
a large numerical grid and calculated 
the single-particle densities for the case of SAE and the converged result of two active 
electrons, CAS$^*(2,21)$, for the lower laser intensity (i).
The results are given in Fig.~\ref{fig:lih-ionization-direction} for both CEPs of the field after the field
is turned off ($t=500$). In the logarithmically
scaled density plot, it becomes apparent for the SAE approximation that ionization is favored if the field points
toward the H atom (``+'', blue). 
Correlation have nearly no effect on 
the single-particle density in this direction  but change the
density emitted in the opposite direction (``-'', red). Here, the small fraction for the SAE case is drastically enhanced for 
the correlated case (dashed line), which in turn results in an increase of the asymmetry parameter $\eta$.

\begin{table}
 \begin{tabular}{lcc}
   \hline
  GAS   & (i) &  (ii)    \\
  \hline
  SAE-v     & 0.12 &  0.27 \\
  CIS-v	    & 0.16 &  0.31 \\
  CAS$^*(2,2)$    & 0.29 &  0.44  \\
  CAS$^*(2,3)$    & 0.22 &  0.39  \\
  CAS$^*(2,6)$    & 0.23 &  0.42  \\
  CAS$^*(2,11)$  & 0.23 &  0.42  \\
  CAS$^*(2,21)$   & 0.23 &  0.42  \\
 \hline
  \end{tabular}
  \begin{tabular}{lcc}
  \hline
  GAS	   & (i) &  (ii)    \\
  \hline
  CIS	   	   & 0.16 &  0.31 \\
  CAS$^*(4,3)$     & 0.30 &  0.48  \\
  CAS$^*(4,4)$     & 0.24 &  0.43  \\
  CAS$^*(4,5)$     & 0.22 &  0.44  \\
 \hline
   & &  \\
   & &  \\
   & &  \\ 
 \end{tabular}
 \caption{Parameter $\eta$ of Eq.~\eqref{eq:eta} for  single-cycle ionization of 
 the 1D 4-electron LiH-like model molecule for peak electrical field strength (i): $F_0=0.025$, (ii): $F_0=0.05$. For $\eta<1$ ($\eta>1$) ionization is more likely when 
 $F_0$ points
 in the direction of Li (H). The left hand side of the table shows GAS results with active 
 valence and fixed core electrons. The right hand side of the table shows GAS results
 where all four electrons are active.}
 \label{tab:ionizationdirection}
\end{table}

\section{Conclusions and outlook}
\label{sec:conclusions}
In this paper, we described and applied the time-dependent generalized-active-space configuration-interaction scheme
to solve the multi-particle time-dependent Schr\"odinger equation. 
The key for the efficient use of TD-GASCI for photoionization of atoms and molecules involving
continua is the use of a partially rotated basis set with HF and pseudo orbitals 
to describe the confined bound state orbitals. Using 1D helium-like and beryllium-like models, we gave a detailed analysis of the convergence 
behavior with respect to the considered orbitals used for the rotation and found that improved
pseudo orbitals based on the $N_{el}-2$ Hartree-Fock problem are well-suited for time-dependent calculations
involving single-electron continua.

We applied the TD-GASCI method to the strong-field ionization of the 1D 4-electron 
LiH-like model 
and found a strong dependence of the observed ionization
yield as a function of the orientation of the molecule with respect to the peak electric field direction and in particular 
on the included level of electron-electron correlation. The e-e interaction increases the ionization
yield in the direction of H. We expect these effects to play also a role in 3D systems.

Although our presented results are for 1D systems, the method is completely general and can
be applied ``as-is'' in arbitrary coordinates. The restriction in dimensionality in this work allowed for a detailed validation of the method through comparison with fully-converged correlated calculations based on the TDSE.
The usability of the similar TD-RASCI approach to single-photon absorption in beryllium and neon in a spherical basis set was 
demonstrated in~\cite{hochstuhl2012,hochstuhl2013}, with the focus on the comparison
with experimental photoionization cross-sections. Generalizations to diatomic molecules in 3D, such as LiH and CO are 
currently in progress based on single-particle orbital expansions in prolate spheroidal coordinates.

\acknowledgments
The authors gratefully thank D. Hochstuhl, H. Miyagi and M. Bonitz for fruitful discussions.
S.B. thanks H. Larsson for helpful comments.
This work was supported by the ERC-StG (Project No. 277767-TDMET), 
the VKR center of excellence, QUSCOPE and the BMBF in the frame of the ``Verbundprojekt FSP 302''.

\appendix
\section{Partial rotation of the single particle basis}
\label{ssec:app-transform}
Let us start with the FE-DVR basis functions $\lbrace | \chi_i \rangle \rbrace$, cf. Eq.~\eqref{eq:fedvr-bridge} and
Eq.~\eqref{eq:fedvr-element} which span the complete simulation region $x\in[0,x_s]$. For simplicity, we consider
only $x \geq 0$. For the interval $[-x_s,x_s]$, the basis can be sorted accordingly.

We partition the basis into a central part $i\in[1,N_c]$ and an outer part $i\in[N_c+1,N_b]$. Because of the orthonormality of the 
FE-DVR functions, we can expand any wave function in the central and the outer part,
\begin{equation}
 |\Psi(t)\rangle = \underbrace{\sum_{i=1}^{N_c} c_i^c(t) |\chi_i \rangle}_{|\Phi^c(t)\rangle} +\underbrace{\sum_{i=1+N_c}^{N_b} c_i^{\textup{o}}(t) | \chi_i\rangle}_{|\Phi^{\textup{o}} (t)\rangle}
 \label{eq:psi-nonrot}
 \end{equation}
Especially, $|\chi_i\rangle \perp |\Phi^\textup{o}\rangle$ for any $i\leq N_c$.

Consider the unitary basis transform 
\begin{equation}
 \boldsymbol{b}=\left(
 \begin{array}{cc}
             \boldsymbol{b}^\textup{c} & \\
                                           & \boldsymbol{1}
  \end{array}
\right ) \;,  
\label{eq:transform-mat}
\end{equation}
which is similar to a rotation of the basis to new basis functions 
\begin{eqnarray}
 |\phi_{\alpha}\rangle&=&\sum_{i=1}^{N_c} \langle \chi_i|\phi_\alpha \rangle |\chi_i\rangle \equiv \sum_{i=1}^{N_c} b_{i\alpha} |\chi_i\rangle \;, \\
 |\chi_i \rangle &=&\sum_{\alpha=1}^{N_c} \langle \phi_\alpha|\chi_i \rangle |\phi_{\alpha}\rangle \equiv \sum_{\alpha=1}^{N_c} b_{\alpha i}^* |\phi_\alpha\rangle \;,
\end{eqnarray}
for $\alpha,i \leq N_c$ and $\chi_i \equiv \phi_i$ else.

The rotated wave function can be analogously to Eq.~\eqref{eq:psi-nonrot} written as
\begin{equation}
 |\Psi_{\textup{rot}}(t)\rangle= |\Phi_{\textup{rot}}^c (t)\rangle  + |\Phi_{\textup{rot}}^o (t)\rangle \;.
\end{equation}
The outer part of the wave function, $|\Phi_{\textup{rot}}^o \rangle$, is thus transformed as
\begin{eqnarray}
 |\Phi^o_\textup{rot} (t)\rangle &=& \sum_{\alpha=N_c+1}^{N_b} \langle \phi_\alpha |\Psi(t)\rangle |\phi_\alpha \rangle \nonumber \\
 &=&\sum_{\alpha=N_c+1}^{N_b} \sum_{i,j} \underbrace{\langle\phi_\alpha |\chi_i \rangle}_{\delta_{\alpha i}} \langle \chi_i | \Psi\rangle(t) \underbrace{\langle \chi_j|\phi_\alpha\rangle}_{\delta_{j\alpha}=\delta_{ij}}|\chi_j\rangle \nonumber\\
 &=&\sum_{i=N_c+1}^{N_b} \langle \chi_i|\Psi(t)\rangle |\chi_i \rangle \nonumber \\
&=& |\Phi^o\rangle \;.
 \end{eqnarray}
Using a transform of type Eq.~\eqref{eq:transform-mat} therefore does not change the outer part of the wave function.
The inner part transforms as
\begin{eqnarray}
 |\Phi^c_\textup{rot} (t)\rangle &=& \sum_{\alpha=1}^{N_c} \langle \phi_\alpha|\Psi(t)\rangle |\phi_\alpha\rangle \nonumber \\
  & =& \sum_{\alpha=1}^{N_c} \sum_{i=1}^{N_c} \langle \phi_\alpha|\chi_i\rangle \langle \chi_i |\Psi(t)\rangle |\phi_\alpha \rangle \nonumber\\
   &=&\sum_{\alpha=1}^{N_c}\sum_{i=1}^{N_c} b_{\alpha i}^* c_i(t) |\phi_\alpha\rangle \;.
\end{eqnarray}
Thus for the central region, $\alpha,i<N_c$, the coefficient vector $c_i$ is transformed to the rotated basis which is
a standard technique in quantum chemistry calculations.

Of course, only the single-particle wave function is invariant under such rotations 
(and so is the full CI many-particle wave function). For truncated CI expansion this is not true, because the truncation error
depends on the accuracy of the single-particle orbitals.
Therefore, the best unitary transformation matrix with the constraint of the boundary at the central and the outer
region has to be found.  Up to now, no straight-forward method to determine this matrix 
for arbitrary time-dependent problems is available, thus the choice
of the transformation matrix $\boldsymbol b$ is guided by physical and mathematical intuition.

\section{Transformation and storage of electron integrals}
\label{sec:app-electron-integrals}
A crucial part for the numerical performance of TD-GASCI calculations is the efficient transformation and storage of the one- and two-electron matrix elements from the FE-DVR to the partially rotated basis.
Extending ideas from~\cite{hochstuhl2014} we evaluate the transformations analytically by exploiting the $\delta$-structure of the FE-DVR matrix elements and the transformation matrix $\boldsymbol{b}$ which
results in fast transformations and offers a strategy for the efficient storage of the transformed integrals.
Similar strategies can be applied for 3D spherical coordinates and prolate spheroidal coordinates.

\subsubsection{Transformation of one-electron integrals}
Let $\langle i|$ and $|j\rangle$ be basis functions from the original FE-DVR set, i.e., with analytically known matrix 
elements $h_{ij}=\langle i|h|j\rangle$ of the single-particle part $h$ of the hamiltonian. 
Let $\langle\alpha|$ and $|\beta\rangle$ denote the rotated mixed basis set,
which can be expanded in the FE-DVR basis as
\begin{eqnarray}
\langle \alpha|&=&\sum_{i=1}^{N_b} b_{\alpha i}^* \langle i| \;\; \textup{and} \;\; |\beta\rangle=\sum_{j=1}^{N_b} b_{\beta j} |j\rangle \; .
\label{eq:mixed-exp}
\end{eqnarray}

The transformation matrix from the FE-DVR basis $|i\rangle$ to the mixed basis $|\alpha\rangle$ is given by Eq.~\eqref{eq:transform-mat},
i.e., for $i,\alpha \in [1,N_c]$, $b_{\alpha,i}$ corresponds to the expansion coefficients of the pseudo orbitals
in the FE-DVR set and for $i,\alpha > N_H$ $b_{\alpha,i}$ is diagonal, $b_{\alpha,i} \equiv \delta_{\alpha,i}$. The latter case corresponds to the situation
$|\beta\rangle = |j\rangle$ and $\langle \alpha | = \langle i|$ outside the central region.

The task is to find the matrix elements of $h$ in the new basis, i.e., $\langle \alpha | h | \beta \rangle\equiv h_{\alpha,\beta}$.
Using Eqs.~\eqref{eq:mixed-exp}, we straightforwardly arrive at
\begin{equation}
 h_{\alpha\beta}=\sum_{i=1}^{N_b}\sum_{j=1}^{N_b} b_{\alpha i}^* \langle i|h|j\rangle b_{\beta j} \; .
\end{equation}
For numerical performance~\cite{bender1972,hochstuhl2014}, at the cost of slightly increased memory consumption,
it is favorable to split this transformation into two parts with a 
temporary matrix $h^{1}$:
\begin{eqnarray}
 h_{i\beta}^{1} & =& \sum_{j=1}^{N_b}b_{\beta j} h_{ij}\;, \nonumber \\
 h_{\alpha \beta} &=&\sum_{i=1}^{N_b} b_{\alpha i}^* h_{i\beta}^1 \; .
\end{eqnarray}
These results can be further simplified by exploiting the diagonal structure 
of the transformation matrix $\boldsymbol{b}$ for $\alpha,i > N_c$,
cf.~Eq.~\eqref{eq:transform-mat},
\begin{eqnarray}
 h_{i\beta}^1&=&\underbrace{\sum_{j=1}^{N_c}b_{\beta j} h_{ij}}_{\textup{if}\;\beta\leq N_c, 0 \; \textup{else}}+ \underbrace{h_{i\beta}}_{\textup{if}\; \beta > N_c, 0 \; \textup{else}}\nonumber \\
 h_{\alpha\beta}&=&\underbrace{\sum_{j=1}^{N_c}b_{\alpha i}^* h_{i\beta}^1}_{\textup{if}\;\beta\leq N_c, 0 \; \textup{else}} + \underbrace{h_{\alpha\beta}^1}_{\textup{if}\;\alpha > N_c, 0 \; \textup{else}} \; .
\end{eqnarray}
Additional straight-forward use of symmetry properties of the FE-DVR matrix elements, 
such as the diagonal or banded structure of the kinetic
and the potential energies, reduces the computational and memory costs further.

\subsubsection{Transformation of two-electron integrals}
For the four-indexed two-electron integrals $w_{ijkl}$, we use a similar approach.
Here, the transformation is split into three parts~\cite{bender1972,hochstuhl2014}. 
Greek letters $\alpha,\beta,\gamma, \delta$  denote transformed indices,
Latin letters $i,j,k,l$ correspond to the untransformed FE-DVR basis:

\begin{enumerate}
 \item $w_{ijkl} \rightarrow w^{(1)}_{ij\gamma\delta}$:
 \begin{equation}
  w^{(1)}_{ij\gamma\delta}=\sum_{k=1}^{N_b} b_{\gamma k}^* \sum_{l=1}^{N_b} b_{\delta l} w_{ijkl}
  \label{eq:2el-1st}
 \end{equation}
 \item $w^{(1)}_{ij\gamma\delta} \rightarrow w^{(2)}_{i\beta\gamma\delta}$:
  \begin{equation}
   w^{(2)}_{i\beta\gamma\delta}=\sum_{j}^{N_b} b_{\beta j}w^{(1)}_{ij\gamma\delta}
   \label{eq:2el-2nd}
  \end{equation}
 \item $w_{i\beta\gamma\delta}^{(2)} \rightarrow w^{(m)}_{\alpha\beta\gamma\delta}$:
  \begin{equation}
    w^{(m)}_{\alpha\beta\gamma\delta}=\sum_{i=1}^{N_b} b^*_{\alpha i}w^{(2)}_{i\beta\gamma\delta}
    \label{eq:2el-3rd}
  \end{equation}
\end{enumerate}
  Due to the special structure of the transformation matrix $b_{\alpha i}$, which is $\delta_{\alpha i}$ for $\alpha,i > N_c$ and the structure of the FE-DVR matrix elements
$w_{ijkl}\propto \delta_{ij}\delta_{kl}$, the above transformations (\ref{eq:2el-1st}-\ref{eq:2el-3rd})
can be simplified ($w_{ik}^F$ denotes the diagonal FE-DVR interaction $w_{ik}^F=w_{ijkl} \delta_{ij}\delta_{kl}$):

\begin{eqnarray}
 w^{(1)}_{ij\gamma\delta} & =&  \begin{cases} \delta_{ij} \sum_{k=1}^{N_c} b_{\gamma k}^* b_{\delta k} w_{ik}^F & \gamma,\delta \leq N_c \\ \delta_{ij} \delta_{\gamma \delta} w_{i\gamma}^F & \gamma, \delta > N_c \\ 0 & \textup{else} \end{cases}
\end{eqnarray}
which can be decomposed into a central part
$w_{i\gamma\delta}^{(1C)}=\sum_{k}^{N_c} b_{\gamma k}^*b_{\delta k} w_{ik}^F$ 
of dimension $N_b \times N_c \times N_c$ and a diagonal part, 
which corresponds to the FE-DVR matrix elements and does not need to be stored.

The second transformation evaluates to
\begin{eqnarray}
  w^{(2)}_{i\beta\gamma\delta} & =& \begin{cases}
  b_{\beta i} w^{(1C)}_{i\gamma\delta} & i,\beta,\gamma,\delta \leq N_c \\
  b_{\beta i} w_{i \gamma}^F \delta_{\gamma,\delta} &  \gamma,\delta > N_c, i,\beta \leq N_c \\  
  b_{\beta i} w_{i\gamma\delta}^{(1C)} \delta_{\beta i} &\gamma,\delta \leq N_c, i,\beta > N_c \\
  w_{i \gamma}^F \delta_{i\beta} \delta_{\gamma\delta}&  i,\beta,\gamma,\delta > N_c 
  \end{cases}
\end{eqnarray}
which gives a central four-indexed part $w^{(2C)}_{i\beta\gamma\delta}$ of dimension $N_c\times N_c \times N_c \times N_c$,
two ``mixed'' parts $w^{(2F1)}$ and $w^{(2F2)}$ of dimension $N_c\times N_c \times N_F$, with $N_F=(N_b-N_c)$ and the ``outer'' diagonal part,
which again corresponds to the FE-DVR matrix elements.

The  transformation can be simplified to
\begin{eqnarray}
 &&w^{(m)}_{\alpha,\beta,\gamma,\delta}  =
\label{eq:2el-ftra}\\
 &&\begin{cases} \sum_{i=1}^{N_c} b_{\alpha i}^* w^{(2C)}_{i\beta\gamma\delta}& \alpha,\beta,\gamma,\delta \leq N_c \\ 
                                                         \delta_{\gamma\delta} \sum_{i=1}^{N_c} b_{\alpha i}^* w^{(2F1)}_{i\beta\gamma} & \alpha,\beta \leq N_c, \gamma,\delta > N_c \\ 
                                                         w^{(2F2)}_{\alpha\gamma\delta}\delta_{\alpha\beta}& \alpha,\beta > N_c, \gamma,\delta \leq N_c \\ 
                                                         \delta_{\alpha \beta} \delta_{\gamma \delta} w_{\alpha\gamma}^F& \alpha,\beta,\gamma,\delta > N_c  \end {cases} \nonumber
\end{eqnarray}

Assuming real-valued orbitals, such as the FE-DVR functions in 1D, the 
symmetry relation for the two-electron integrals, $w_{\alpha\beta,\gamma\delta}=w_{\gamma\delta,\alpha\beta}$,
reduces the storage requirements to $w^C_{\alpha\beta,\gamma\delta}$ [first row of Eq.~\eqref{eq:2el-ftra}]
and either $w^{F1}$ or $w^{F2}$ [second or third row of Eq.~\eqref{eq:2el-ftra}].
Thus in total two arrays have to be stored. The central array $w^{H}_{\alpha,\beta,\gamma\,\delta}$ for $\alpha,\beta,\gamma,\delta \leq N_H$ 
of dimension $N_c\times N_c \times N_c \times N_c$ and
one mixed, three-indexed, array $w^{F1}_{\alpha,\gamma,\delta}$ or $w^{F2}_{\alpha,\beta,\gamma}$ 
of dimension $N_F \times N_c \times N_c$.
This allows for an efficient storage scheme of the two-electron integrals in the mixed basis set approach and with that for the application of GASCI to large extended systems (e.g.~photoionization)
without approximation of the interaction matrix elements.

\end{document}